\documentclass[%
reprint,
superscriptaddress,
nofootinbib,
amsmath,amssymb,
 aps,physrev,
 longbibliography,
]{revtex4-2}
\usepackage[utf8]{inputenc}
\usepackage[T1]{fontenc}
\usepackage{bm}
\usepackage{amsmath}
\usepackage{amssymb}
\usepackage{amsfonts}
\usepackage[switch]{lineno}

\usepackage{mathtools}

\usepackage{placeins}

\usepackage{xcolor} %

\usepackage{graphicx}%
\usepackage{dcolumn}%
\usepackage{bm}%
\usepackage{siunitx}

\newcommand{\Eqref}[1]{Eq.~\eqref{#1}}
\usepackage[english]{babel}
\usepackage{textcomp}
\usepackage[toc,page]{appendix}
\usepackage{comment}
\usepackage{indentfirst}
\usepackage[separate-uncertainty=true]{siunitx}[=v2]
\usepackage{color}
\usepackage[colorlinks=true,allcolors=blue]{hyperref}
\usepackage[normalem]{ulem}
\usepackage{soul}
\usepackage{setspace}
\usepackage{array}
\usepackage{booktabs}
\usepackage{tabularx}
\usepackage{grffile}
\usepackage{float}
\usepackage{lmodern}%

\DeclareSIUnit[number-unit-product = {}]{\pixel}{pixel}
\DeclareSIUnit[number-unit-product = {}]{\min}{min}
\usepackage{comment}
\graphicspath{{./Figures/}}

\newcommand{\swb}[1]{\textcolor{broMaterial }} %

\newcommand*\diff{\mathop{}\!\mathrm{d}} %

\newcommand{\ladhyx}{Laboratoire  d'Hydrodynamique (LadHyX), CNRS, Ecole Polytechnique, Institut Polytechnique de Paris, 91120 Palaiseau, France}
\newcommand{\tum}{Center for Protein Assemblies and Department of Bioscience, School of Natural Sciences, Technische Universität München, 85748 Garching, Germany}
\newcommand{\lmu}{Faculty of Physics and Center for NanoScience, Ludwig-Maximilians-Universität München, 80539 Munich, Germany}
\newcommand{\mpi}{Max Planck School Matter to Life, Max Planck Institute for Medical Research, Heidelberg, 69120, Germany}

\begin{document}

\title{Self-organized homogenization of flow networks}

\author{Julien Bouvard}
\thanks{These authors contributed equally to this work.}
\affiliation{\ladhyx}

\author{Swarnavo Basu}
\thanks{These authors contributed equally to this work.}
\affiliation{\tum}

\author{Charlott Leu}
\affiliation{\lmu}

\author{Onurcan Bektas}
\affiliation{\tum}
\affiliation{\lmu}
\affiliation{\mpi}

\author{Joachim O.~Rädler}
\affiliation{\lmu}

\author{Gabriel Amselem}
\email[Contact author: ]{gabriel.amselem@polytechnique.edu}
\affiliation{\ladhyx}

\author{Karen Alim}
\email[Contact author: ]{k.alim@tum.de}
\affiliation{\tum}

\date{\today}

\begin{abstract}
From the vasculature of animals to the porous media making up batteries, the core task of flow networks is to transport solutes and perfuse all cells or media equally with resources. Yet, living flow networks have a key advantage over porous media: they are adaptive and self-organize their geometry for homogeneous perfusion throughout the network. Here, we show that also artificial flow networks can self-organize toward homogeneous perfusion by the versatile adaption of controlled erosion. Flowing a pulse of cleaving enzyme through a network patterned into an erodible hydrogel, with initial channels disparate in width, we observe a homogenization in channel resistances. Experimental observations are matched with numerical simulations of the diffusion-advection-sorption dynamics of an eroding enzyme within a network. Analyzing transport dynamics theoretically, we show that homogenization only occurs if the pulse of the eroding enzyme lasts longer than the time it takes any channel to equilibrate to the pulse concentration. The equilibration time scale derived analytically is in agreement with simulations. Lastly, we show both numerically and experimentally that erosion leads to the homogenization of complex networks containing loops. Erosion being an omnipresent reaction, our results pave the way for a very versatile self-organized increase in the performance of porous media.
\end{abstract}

\flushbottom
\maketitle

\section{Introduction}
Whether in engineered or natural systems, fluid flows perfuse disordered porous media, from packed-bed reactors, filters, bio-engineered tissues and fuel cells, to rocks, geological sediments, and irrigated soils~\cite{Stankey.2024,sun2019hierarchical,tien1979advances,bowen1995theoretical,beven1982macropores}.  The disordered structure of their networks' channels results in heterogeneous flows~\cite{Lebon.1996,Kutsovsky.1996,Shattuck.1997, Maier.1998, Sederman.2001, Moroni.2001, Datta.2013, alim2017local,Matyka.2016}, diminishing perfusion in network channels of high resistance~\cite{Meigel.2022}. The consequences of the lack in perfusion are dramatic, reducing the efficiency of packed-bed reactors, filters and fuel cells and the survival of cells in bio-engineered tissues \cite{Sexton.2025}. 

In contrast, in biological flow networks such as fungal mycelia, slime mold veins, or our own microvasculature~\cite{isogai2001vascular,boddy2009saprotrophic,tero2010rules}, network channels adapt such that network geometry is self-organized to transport solutes throughout the network, perfusing all parts of the organism with nutrients and signaling molecules~\cite{chang2019microvascular,Chang.20175gc,Kirkegaard.2020,simonin2012physiological,alim2017mechanism,Liese.2021,Kramer.2023, Karschau.2020}. 
Such perfusion is best achieved when the velocity of the fluid is homogeneous in all channels of a network like in intermeshed channels of equal resistance~\cite{chang2019microvascular,Chang.20175gc,Kirkegaard.2020,Liese.2021,Kramer.2023,meigel2018flow,Qi.2021}. Such uniformity, though, is contrary to the natural tendency to generate disorder and thus inhomogeneities~\cite{alim2017local}. The key to living flow networks' self-organization is the local adaptation of channels to local flow rate~\cite{Murray.1926krhf,marbach2023vein,zareei_temporal_2022} or local solute concentrations~\cite{chang2019microvascular,gavrilchenko2021distribution,meigel2018flow,kramar2021encoding}. Such local adaptation is sufficient to drive global self-organization of the flow networks as flows are globally coupled due to conservation of fluid volume: changing locally a channel width changes the local resistance to flow, which overall redistributes fluid throughout the entire network. Thus, even if inanimate flow networks cannot rise to the complexity of local adaptation as living networks, any local adaptation will hijack the global coupling of flows.

Engineered flow networks operating at dimensions similar to those of living networks are readily reproduced using microfluidic tools~\cite{wong2012microfluidic,sebastian2018microfluidics,fenech2019microfluidic}. Integrating responsive materials renders microfluidic chips adaptive to local solutes controlled by transport within fluid flows~\cite{beebe2000functional,eddington2004flow}. In particular, the swelling and contraction of hydrogels by local solute concentration, or temperature~\cite{richter2003electronically,d2018microfluidic}, has so far only been employed to operate switches, closing or opening a channel~\cite{beebe2000functional,baldi2002hydrogel,liu2002fabrication,goy2019microfluidics,paratore2022reconfigurable,na2022hydrogel}.
Yet, carving entire networks of channels into adaptive hydrogels opens the potential for self-organizing flow networks.
\begin{figure*}[t]
\includegraphics[width = 1.0\textwidth]{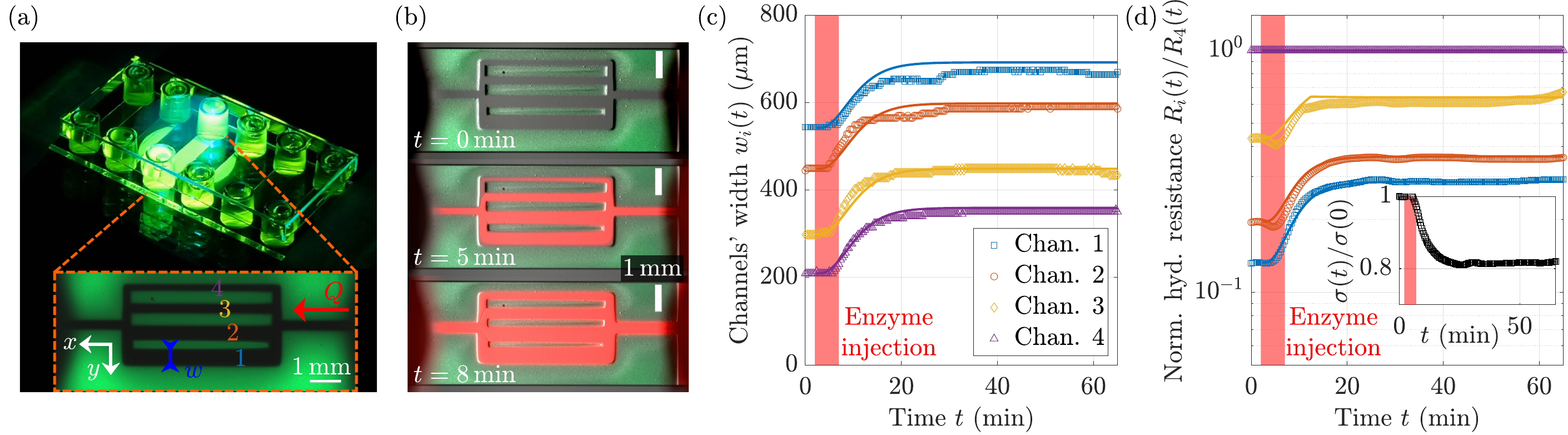}
  \caption{\textbf{Erosion by the MMP-1 enzyme homogenizes the resistances of parallel channels whose walls are made of the hydrogel PEG-NB, cross-linked with a cleavable peptide sequence.} (a) Experimental setup: glass slide with six microfluidic chambers. Each chamber contains a different microfluidic network, whose hydrogel walls are made of PEG-NB cross-linked with a cleavable peptide sequence. Inset: four-channel network. (b) Time-lapse of the erosion of a four-channel network, whose hydraulic resistances are initially imbalanced. The images, taken at $t=0$ (top), $t=\SI{5}{min}$ (middle), and $t=\SI{8}{min}$ (bottom), show the hydrogel (light green) being eroded by the enzyme (red). Scale bar: \SI{1}{\mm}. (c) Time-evolution of the widths $w_i(t)$ of the four channels. The injection of the enzyme upstream of the network occurs from $t=2$\,min to $t=7$\,min (light red patch). Points: experimental data. Lines: numerical simulations. (d) Homogenization of the channels' hydraulic resistances $R_i(t)$ as the channel walls are getting eroded: the hydraulic resistances $R_i(t)/R_4(t)$ normalized by the narrowest channel's resistance homogenize as they increase over time. Channels are numbered from the largest to the smallest width. Points: experimental data. Lines: numerical simulations. Inset: normalized standard deviation $\sigma(t)/\sigma(0)$ of the normalized hydraulic resistances $R_i(t)/R_4(t)$. Similarly to the hydraulic resistances, the flow rates also homogenize, see Supp.~Fig.~11~\cite{supp}.}
\label{fig:1}
\end{figure*}

Here\nocite{weber2009effects}, we combine experiments, theory, and numerical simulations to develop adaptive inanimate flow networks capable of self-organization for homogeneous flow independent of their initial network geometry by controlled erosion. We build microfluidic channel networks out of an erodible hydrogel and let the device geometry progressively adapt as pulses of erosive chemicals flow into the chip. When the chemical diffuses into the hydrogel walls, these erode, which changes the hydraulic resistances of the channels and, thus, the overall flow pattern in the network. Experiments are successfully described by numerical solutions of the advection-diffusion-sorption dynamics of the flow-transported enzyme within the network. The theoretical description predicts homogenization by erosion to generally hold for any flow network as long as the pulse length is longer than the time for equilibration of enzyme concentration in any channel. We analytically derive the enzyme equilibration time scale as an easy-to-apply rule to achieve homogenization by controlled erosion. Successful homogenization of complex loopy microfluidic networks confirms the predicted general applicability. Our work paves the way for the design of a novel class of adaptive flow devices and versatile optimization of porous media performance by such a simple adaptation as erosion.
\section{Results}
\subsection{Erosion homogenizes resistances of parallel channels}
We begin by considering an imbalanced network of four parallel channels of identical height ($H=\SI{400}{\um}$) and length ($L=\SI{4250}{\um}$), but whose widths $w_i$ vary from $210$ to $\SI{545}{\um}$, %
see Fig.~\ref{fig:1}a. The top and bottom of the channels are sealed with plastic, while their side walls consist of the hydrogel PEG-NB, cross-linked with a cleavable peptide sequence (see Methods). To probe the impact of controlled erosion, we inject a $t_e=$ \SI{5}{min} long pulse of the eroding enzyme MMP-1 at a concentration of $c_0=\SI{3e-5}{\mol\per\liter}$ and a flow rate of $Q=\SI{20}{\uL\per\min}$, followed by the injection of Phosphate Buffer Saline (PBS) for \SI{60}{min}. The boundaries of the channel wall and the dynamics of the enzyme are monitored under a fluorescence microscope. The walls fluoresce in green and the enzyme solute in red due to the incorporation of a green fluorescent dye into the hydrogel mesh, and the mixing of the MMP-1 enzyme with a red fluorescent dye of the same molecular weight, Texas-red dextran (see Methods). Snapshots of typical experiments are shown in Fig.~\ref{fig:1}b for a four-channel network and in Supp.~Fig.~2a~\cite{supp} for a single channel one (see also Movie~1 in the Supplemental Material~\cite{supp}).

When the eroding solution is injected, the enzyme concentration increases both in the channels and, due to the enzyme's diffusivity, within the hydrogel walls. In the presence of the enzyme, the peptide sequence cross-linking the hydrogel is cleaved, leading to the erosion of the channel walls and, thus, to an increase in channel widths (see Fig.~\ref{fig:1}c). Flowing PBS into the network flushes away the eroding solution and gradually slows down the erosion.

\begin{figure*}[!ht]
\centering
  \includegraphics[width = 1.0\textwidth]{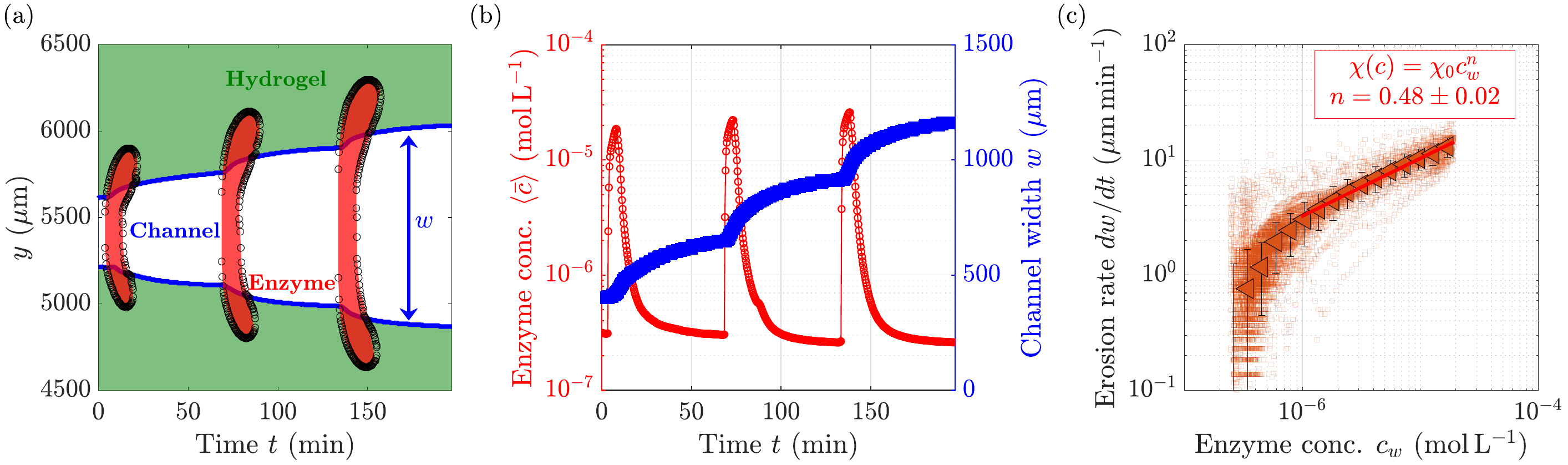}
  \caption{\textbf{Wall erosion follows Michaelis-Menten kinetics.} (a) Spatio-temporal evolution of the channel width $w$ of a single channel, see Movie~1 in the Supplemental Material~\cite{supp}, at a fixed position along the channel. The hydrogel (light green) is eroded by the enzyme (red, $c\geq\SI{1.5e-6}{\mol\per\liter}$) diffusing into the hydrogel and back into the channel after the pulse passed. The hydrogel walls, i.e.~the boundaries between the channel and the hydrogel, are highlighted in blue. Three pulses of enzyme of duration $t_e=5$\,min are inter-spaced by 60\,min of PBS flowing into the network. (b) Time-evolution of the channel width $w$ (blue) as pulses of the enzyme are flowed in. The concentration of the enzyme in the channel (red) is averaged along the cross-section of the channel in $y$ and along the channel in $x$, $\langle\bar{c}\rangle$. %
  (c) Erosion rate $dw/dt$ of the hydrogel, as a function of the local enzyme concentration $c_w$ within \SI{50}{\um} of the wall, follows Michaelis-Menten kinetics. Squares: raw data. Triangles: averaged data. Error bars show the standard deviation. Line: power-law fit of the data. %
  Data over 10 experiments with 10 different channel geometries but keeping the PBS concentration and channel heights constant. The kinetic model is consistent over variations of height and buffer concentration in altered gels, see Supp.~Fig.~2c~\cite{supp}.} %
  \label{fig:2}
\end{figure*}
To address how erosion affects the imbalance of the channels, we calculate the evolution of the hydraulic resistances of the four parallel channels from measurements of the channels' widths as a function of time. The hydraulic resistance for a rectangular channel of length $L$, width $w$ and height $H>w$, is given by
\begin{equation}
\label{eq:resistance}
    R = \dfrac{12\eta L}{H w^3} \left(1-0.63\dfrac{w}{H}\right)^{-1},
\end{equation}   
where $\eta$ is the fluid's viscosity~\cite{Bruus.2011}. Normalizing all channel resistances by the resistance of the narrowest channel in the network at any point in time reveals that the network homogenizes in resistances with time. The smallest resistance in the network is initially $\approx 10$ times smaller than the largest one; after one pulse of eroding enzyme, the smallest and the largest resistances are within a factor $\approx 3$, see Fig.~\ref{fig:1}d. The normalized standard deviation $\sigma(t)/\sigma(0)$ of the normalized resistances $R_i(t)/R_4(t)$ decreases from 1 to 0.8 thanks to erosion, see inset of Fig.~\ref{fig:1}d. Likewise, the normalized flow rates homogenize, see Supp.~Fig.~11~\cite{supp}, with the normalized standard deviation halving from 1 to 0.54, see inset of Supp.~Fig.~11~\cite{supp}. Thus, we find flow network adaptation by controlled erosion dynamics to homogenize both channel resistances and flow rates.

\subsection{Interplay of simulation and data yield quantitative prediction of erosion dynamics }
To quantitatively understand the erosion dynamics, we follow the erosion in a device reduced in network complexity to a single PEG-NB channel, see Movie~1 in the Supplemental Material~\cite{supp}. We measure the channel width and enzyme concentration across the channel and at the hydrogel wall over three subsequent erosion enzyme pulses of duration $t_e=\SI{5}{min}$, each followed by $\SI{60}{min}$ of PBS solution to flush out the enzyme, see kymograph of the channel width in Fig.~\ref{fig:2}a. In the kymograph, diffusion of the enzyme from the channel into the hydrogel, and then from the hydrogel into the channel, leads to zones of high enzyme concentration in the hydrogel with a curved boundary (see red zones in Fig.~\ref{fig:2}a). The red zones correspond to regions of medium to high enzyme concentrations, $c>\SI{1.5e-6}{\g\per\mol}$, i.e.~regions where the enzyme concentration is diluted less than 20 times compared to the injected concentration $c_0=\SI{3e-5}{\mol\per\liter}$. For further analysis, both the enzyme concentration $c$ and the channel width $w$ are averaged over the entire channel length. 
Upon inflow of a pulse of the MMP-1 enzyme, the channel starts to erode, and its width increases at an average rate of $dw/dt \approx \SI{5}{\um\per\min}$, see Supp. Fig.~2b~\cite{supp}. When, after a time $t_e=\SI{5}{min}$, PBS is flowed in to replace the enzyme, the enzyme concentration in the channel decays to zero. During this time, enzyme molecules that had diffused into the hydrogel walls now diffuse back into the main channel and are advected away by the flow of PBS. The hydrogel walls keep eroding, albeit at a slower and slower rate, see Fig.~\ref{fig:2}a,b and Supp.~Fig.~2b~\cite{supp}. 

To understand how the enzyme controls the channel erosion rate $dw/dt$, we compute the wall erosion rate at all locations along the channel wall and measure the enzyme concentration locally within \SI{50}{\um} of the channel wall, which we denote $c_w$ to distinguish it from the concentration within the channel. The higher the local enzyme concentration $c_w$, the larger the erosion rate, see Fig.~\ref{fig:2}c. For concentrations between $c_w=\SI{1e-6}{\mol\per\liter}$ and $c_w=\SI{2e-5}{\mol\per\liter}$, the erosion rate increases as $dw/dt\sim\sqrt{c_w}$, with typical values around $dw/dt\approx\SI{5}{\um\per\minute}$ (see Supp.~Fig.~2b~\cite{supp}).
We derive the erosion rates' scaling law from the enzyme's diffusion into the hydrogel, coupled to Michaelis-Menten kinetics for the enzymatic reaction. Call $A$ the concentration of peptide crosslinker and $c_w$ the enzyme concentration. The erosion rate of the cross-linkers is given by $dA/dt = -k_{\rm cat} c_w A/(K_M + A)$, where $k_{\rm cat}$ is the kinetic erosion constant in $\si{\per\s}$ and $K_M$ is the Michaelis constant in $\si{\mol\per\liter}$. In our experiments, the initial concentration of crosslinker $A_0=\SI{20}{\mmol\per\liter}$ is much larger than typical values reported for the Michaelis constant of the hydrolysis of similar peptide chains by MMP-1, $K_M\approx \SI{0.5}{\mmol\per\liter}$~\cite{lutolf2003synthetic,nagase1996human}. Therefore, our characteristic time scale of the reaction is $T\sim A_0/(k_{\rm cat}c_w)$. During this time, the enzyme has diffused a distance $Y\sim\sqrt{DT}$ into the hydrogel, with $D$ the diffusion coefficient of the enzyme in the hydrogel. The erosion rate $dw/dt$ is then proportional to
\begin{equation}
\label{eq:scaling}
\frac{dw}{dt}\sim\dfrac{Y}{T}\sim\sqrt{\dfrac{Dk_\mathrm{cat}c_w}{A_0}}=\chi_0\sqrt{c_w},
\end{equation}
recovering the square-root relationship. 
Note that the only unknown in Eq.~\eqref{eq:scaling} is the erosion constant $k_\mathrm{cat}$. Fitting the scaling law to our data, we extract $k_{\rm cat} \approx \SI{2.1\pm0.7}{\per\s}$, respectively %
$\chi_0=\SI{3.3\pm0.6e3}{\um\per\min\mol\tothe{-1/2}\liter\tothe{1/2}}$, in good agreement with the literature~\cite{nagase1996human,lutolf2003synthetic,patterson2010enhanced}, see Supp.~Mat.~3~\cite{supp}. The model is consistent over variations of height and buffer concentration in altered gels, see Supp.~Fig.~2c~\cite{supp}.%

With a quantitative description of the erosion dynamics as a function of the enzyme concentration at hand, we next turn toward the dynamics of the enzyme within a channel of constant width $w$. 
We approximate the flow profile as parabolic along the width $w$. The cross-sectional flow velocity of the Poiseuille flow $\bar{U}$ under these assumptions is $\bar{U}=Q/wH$, where $Q$ denotes the flow rate. To optimize the numerical integration of enzyme dynamics, we reduce the dimension of the problem to the cross-sectional average concentration $\bar{c}(x,t)$ evolving only along the channel's longitudinal axis $x$, commonly known as Taylor dispersion~\cite{Taylor.1953,Aris.1956}. We describe the enzyme diffusion into and out of the hydrogel with enzyme absorption and desorption kinetics into the wall, thereby neglecting the spatial variation in enzyme concentration within the channel wall, which is a good assumption for surface erosion-dominated kinetics of hydrogels~\cite{Kharkar.2013}. For a solute getting absorbed in channel walls with rate $K_a$ and desorbed back into the channel stream with rate $K_d$, the series expansion of Zhang \emph{et al.}~\cite{zhang_transient_2017} captures the cross-sectional average solute $\bar{c}(x,t)$ and absorbed solute dynamics $c_w(x,t)$ not limited by Péclet number, 
\begin{subequations}
\label{eq:finaladvdiff}
\begin{align} 
    \frac{\partial \bar{c}}{\partial t}&=-\frac{\alpha K_a}{w} \bar{c}+\frac{\alpha K_d}{w} C_w-U^{\mathrm{eff}} \frac{\partial \bar{c}}{\partial x}+D^{\mathrm{eff}} \frac{\partial^2 \bar{c}}{\partial x^2}, \label{eq:cbar}\\
    \frac{\partial C_w}{\partial t} &= K_a \bar{c} - K_d C_w \label{eq:gamma},
\end{align}
\end{subequations}
where we expand up to second order for the effective flow velocity $U^{\mathrm{eff}}(t)$ and effective diffusivity $D^{\mathrm{eff}}(t)$, which are functions of $K_a$, $K_d$, the enzyme's molecular diffusivity $D$, and the cross-sectional flow velocity $\bar{U}$. For the absorption kinetics, we remain at first order to extract  the surface concentration, $C_w=\Tilde{l}c_w$ directly with $\Tilde{l}=\SI{1}{\milli\m}$, and $\alpha=10^{-3}$ a non-dimensional parameter encoding the wall thickness, see Supp.~Mat.~10~\cite{supp}.  The advection-diffusion-sorption dynamics are therefore specified by three dimensionless parameters - the Péclet number defined as $\mathrm{Pe}=\bar{U}w/D$, and two Damköhler numbers defined as $k_a=K_aw/D$ and $k_d=K_dw^2/D$. Note, that due to the desorption dynamics, effective flow velocity and effective diffusivity are time dependent variables in contrast to the static dimensionless parameters. We use $D=\SI{30}{\square\um\per\s}$ for the enzyme's molecular diffusivity, based on experimental measurements (see Supp.~Fig.~1~\cite{supp}), which are in agreement with the literature~\cite{gaspers1994enzymes}.
The absorption rate $K_a$ and desorption rate $K_d$ are extracted by fitting the experimental channel wall concentration to Eq.~\eqref{eq:gamma}. The rates are found to be $K_a=\SI{8.4}{\um\per\s}$ and $K_d=\SI{11.4}{\per\s}$ (see Supp.~Fig.~5~\cite{supp}). 
Invoking our erosion law \Eqref{eq:scaling}, the channel width $w(x,t)$ is predicted from $c_w(x,t)$ by
\begin{equation}
     w(x,t) = w_0 + \chi_0 \int_0^{t} \sqrt{c_w (x,\tilde{t})} \diff \tilde{t},
     \label{eq:widthdyn}
\end{equation}
where $w_0$ is the initial width of the channel. %
This numerical model is now our stepping stone in solving the erosion dynamics of channel networks and understanding the mechanism driving homogenization.

\begin{figure*}[t]
\centering
    \includegraphics[width = 1.0\textwidth]{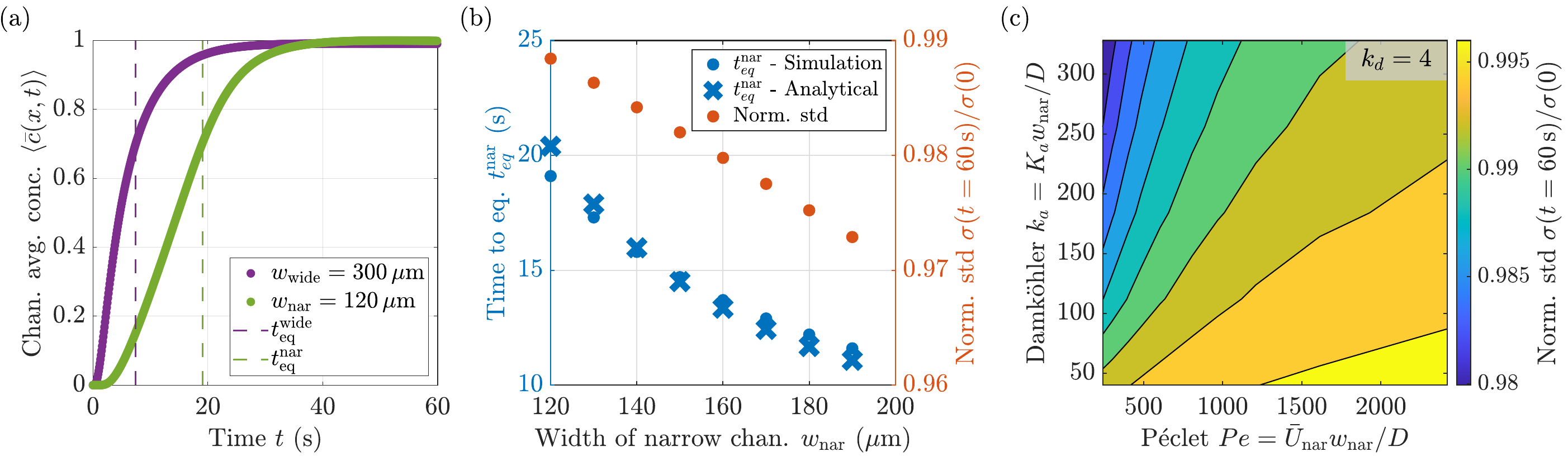}  
    \caption{\textbf{High absorption, low desorption and reducing Péclet number improve homogenization.} (a) Normalized enzyme concentration averaged over the channel length $\left<\bar{c}(x,t)\right>$ for two parallel channels shows a slower rise in the narrow channel (green) relative to the wide channel (purple). Channels are of equal length $L=\SI{4.5}{mm}$. The time scale to reach the upstream equilibrium concentration $t_\mathrm{eq}$ is extracted as the time-point of $\left<\bar{c}(x,t)\right>$ crossing 0.7. (b) Low values of $t_{\mathrm{eq}}$ in the narrow channel, $t_{\mathrm{eq}}^{\mathrm{nar}}$, correspond to better homogenization - the shorter it takes for the concentration in the narrow channel to reach the upstream concentration, the more the channel erodes, leading to better homogenization of the hydraulic resistances reflected by a greater decrease in  $\sigma(t=\SI{60}{\s})/\sigma(0)$. Numerically determined $t_{\mathrm{eq}}$ (blue dots) obtained from (a) for different values of $w_{\mathrm{nar}}$ with a constant value for $w_{\mathrm{wide}}=$ \SI{300}{\um} match the corresponding analytically calculated $t_{\mathrm{eq}}$ from Eq.~\eqref{eq:t_eq} (crosses). The simulation parameters are: 
    diffusivity $D=\SI{30}{\square\um\per\s}$, absorption rate $K_a=\SI{20}{\um\per\s}$ and desorption rate $K_d=\SI{30}{\per\s}$.
    (c) Homogenization $\sigma(t=\SI{60}{\s})/\sigma(0)$ (color coded) by erosion is best for high absorption, low desorption. Also, reducing the Péclet number increases homogenization for $\mathrm{Pe}>1$ considered here. Here, we only show the plane with $k_d=K_dw^2_{\rm nar}/D=4$ for clarity; see Supp.~Fig.~13a~\cite{supp} and Movie 6 in the Supplementary Material for the full 3D phase plot~\cite{supp}. $D$, $K_a$ and $K_d$ varied for $w_{\mathrm{nar}}=$ \SI{200}{\um} and  $w_{\mathrm{wide}}=\SI{300}{\um}$. For all plots, homogenization dynamics employ a channel height $H=\SI{1}{mm}$, enzyme concentration $c_0=\SI{3e-5}{\mol\per\liter}$, erosion constant $\chi_0=\SI{3.3\pm0.6e3}{\um\per\min\mol\tothe{-1/2}\liter\tothe{1/2}}$, and inflow rate $Q=\SI{6}{\uL\per\min}$. Note that the pulse duration $t_e$ was set to $t_e=\SI{1}{min}$ for numerical efficiency. Increasing the pulse length increases the erosion, which results in a steeper drop in the homogenization metric $\sigma(t)/\sigma(0)$, indicating better homogenization, see Supp.~Fig.~13c~\cite{supp}.} %
  \label{fig:theory}
\end{figure*}

\subsection{Homogenization by erosion is robust for pulse concentration equilibration faster than pulse duration}
To model erosion in channel networks, we map experimental network designs to a network skeleton consisting of network edges connected at network nodes. Each network edge represents a channel of constant width, whose hydraulic resistance is given according to channel geometry following \Eqref{eq:resistance}. Imposing inflow and matching outflow at network inlet and outlet, respectively, flow rates in all channels and, thus, corresponding flow velocities follow from imposing conservation of fluid volume aka Kirchhoff's law at all network nodes~\cite{Meigel.2022}. We then solve the advection-diffusion-sorption dynamics, Eqs.~\eqref{eq:finaladvdiff}, in each channel of the network (see Methods). Numerically solving for the enzyme dynamics for our initial four-channel experimental parameters of $w_1=\SI{540}{\um}$, %
$w_2=\SI{450}{\um}$, %
$w_3=\SI{300}{\um}$, $w_4=\SI{210}{\um}$, $H=\SI{400}{\um}$ and $Q=\SI{20}{\uL\per\min}$ we obtain the cross-sectional average enzyme concentration in the channel $\bar{c}(x,t)$, the absorbed enzyme concentration $c_w(x,t)$ and subsequently the channel width $w(t)$, see Fig.~\ref{fig:1}c,d, in excellent agreement with experimental observations. In particular, simulations successfully predict our hallmark of homogenization, the decrease in the normalized standard deviation of the normalized hydraulic resistances $\sigma(t)/\sigma(0)$, quantitatively. Notably, homogenization is achieved although all channels erode almost the same amount: The change in channel width over the course of a single pulse $\Delta w_i$ in the imbalanced four channels is within \SI{6}{\%} for all channels.

To investigate analytically how the amount of wall erosion affects homogenization, we consider the resistance ratio of, for simplicity, two parallel channels of equal height and length but one narrow and one wide in width. As the hydraulic resistance $R$ of a channel depends non-linearly on the channel width: $R\sim w^{-3}$, see \Eqref{eq:resistance}, an initially small resistance ratio of a wide channel of low resistance $R_{\rm wide}$ to a narrow channel of high resistance $R_{\rm nar}$ is bound to increase if both channels erode equally $\Delta w_{\rm wide}=\Delta w_{\rm nar}$,
\begin{equation}
\label{eq:homo}
\frac{R_{\rm wide}(w_{\rm wide}+\Delta w)}{R_{\rm nar}(w_{\rm nar}+\Delta w)} = \frac{R_{\rm wide}(w_{\rm wide})}{R_{\rm nar}(w_{\rm nar})}\frac{\left(1+\frac{\Delta w_{\rm nar}}{w_{\rm nar}}\right)^3}{\left(1+\frac{\Delta w_{\rm wide}}{w_{\rm wide}}\right)^3}.
\end{equation}
In fact, homogenization will always take place as long as the fraction of the two cubic terms in \Eqref{eq:homo} is bigger than one. Here, we remind ourselves that wall erosion is determined by Michaelis-Menten kinetics, captured by \Eqref{eq:widthdyn}. Thus, homogenization by erosion is tied to the variation of enzyme absorbed into the channel walls $c_w$ and, consequently, enzyme distributed within channels across a channel network. 

\begin{figure}[t]
\centering
    \includegraphics[width = 0.48\textwidth]{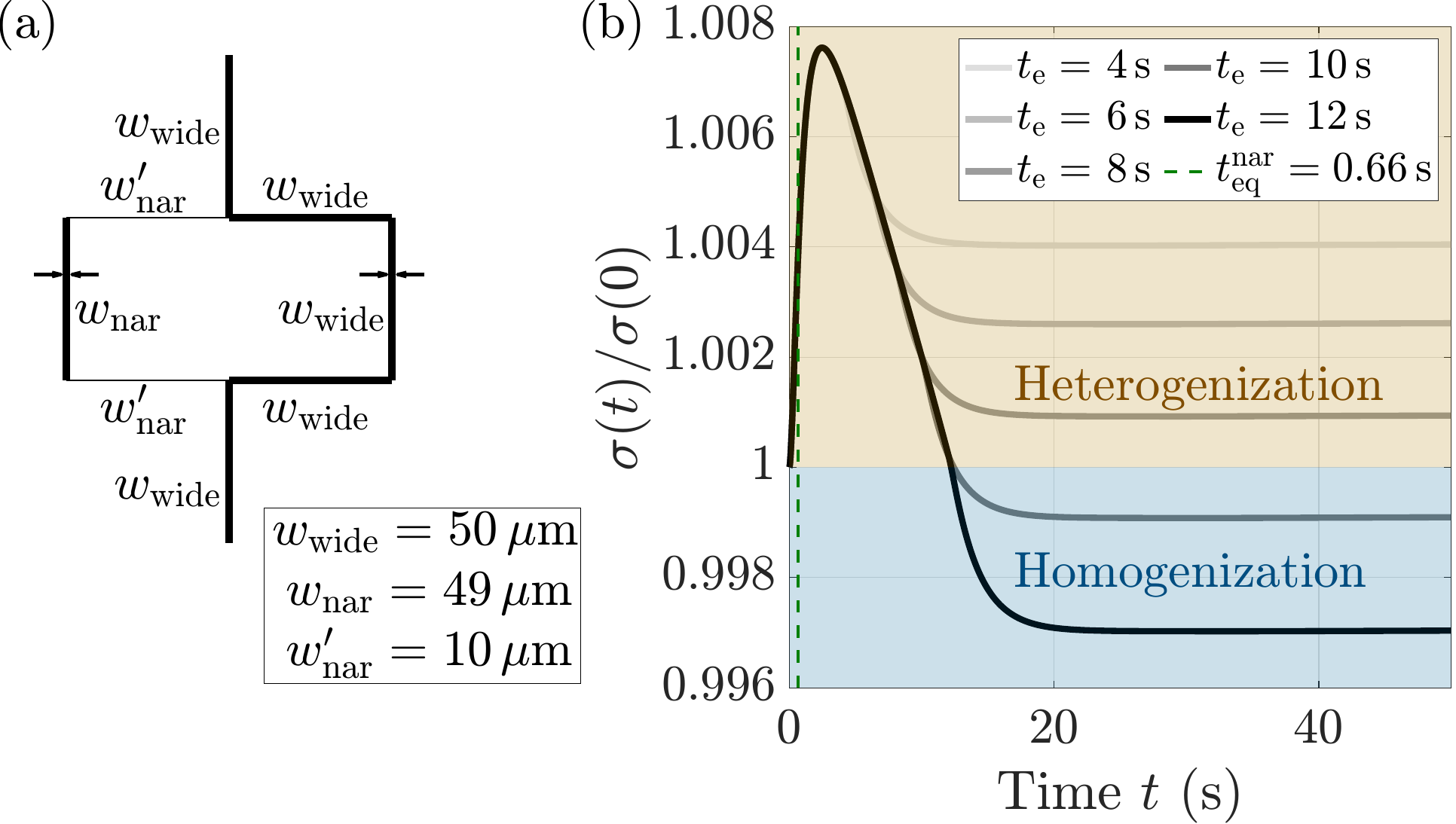}  
    \caption{\textbf{Transition from heterogenization to homogenization at increasing erosion pulse length.} (a) Two parallel channel design where parallel channels are now partitioned into three sections of varying width. The disparity in width in the entry zone of the parallel channels $w'_{\text{nar}}\ll w_{\text{wide}}$ creates a much longer time to equilibrate in the narrower channel than in the wider channel. The small difference in width in the center section $w_{\text{nar}}<w_{\text{wide}}$ reduces the required heterogeneity in channel width erosion for heterogenization, $\Delta w_{\text{nar}}/\Delta w_{\text{wide}}<w_{\text{nar}}/w_{\text{wide}}$. (b) The normalized standard deviation $\sigma(t)/\sigma(0)$ of the normalized hydraulic resistances $R_{\text{nar}}(t)/R_{\text{wide}}(t)$ of the center sections increases as the erosion pulse passing is shorter than 10\,s but decreases at longer pulse length, much larger than the time to equilibrate in the narrow parallel channel $t_{\text{eq}}=\SI{0.66}{\s}$. The simulation parameters are: inflow rate $Q=\SI{60}{\uL\per\min}$, 
    diffusivity $D=\SI{3}{\square\um\per\s}$, absorption rate $K_a=\SI{20}{\um\per\s}$ and desorption rate $K_d=\SI{e3}{\per\s}$ and channel height $H=\SI{200}{\um}$.} %
  \label{fig:transition}
\end{figure}

Therefore, we next address how the dynamics of enzyme distribution over a junction within a two-channel network affect homogenization. Mapping out the evolution of the normalized channel-averaged enzyme concentration $\left<\bar{c}(x,t)\right>=\int_0^L\bar{c}(x,t)\diff{x}/(C_0L)$ over the course of a \SI{1}{min} pulse in two parallel channels disparate in width shows that both channels eventually reach the upstream enzyme concentration. Yet, the rise in enzyme concentration is much slower in the narrower channel compared to the wider channel, see Fig.~\ref{fig:theory}a. We assess the impact of these disparate enzyme dynamics on homogenization by scoring the normalized standard deviation of the normalized hydraulic resistance $\sigma(t)/\sigma(0)$, while systematically varying the width of the narrower channel but keeping the width of the wide channel constant, see Fig.~\ref{fig:theory}b. Over time $\sigma(t)/\sigma(0)$ decreases from 1 when the resistances homogenize. Thus, a lower final value indicates better homogenization. We observe that homogenization is less and less successful if the narrower channel is initially considerably narrower than the wide channel. We hypothesize that a narrower channel results in a longer time to equilibrate the inflowing enzyme concentration to the upstream enzyme concentration, such that there is less erosion of the narrower channel compared to the wide channel and, therefore, less homogenization. To test our hypothesis, we quantify the timescale to equilibrate to the upstream enzyme concentration $t_{\mathrm{eq}}$ as the time at which the concentration reached \SI{70}{\%} of the upstream enzyme concentration, see Fig.~\ref{fig:theory}a. Indeed, long equilibration timescales accurately predict low homogenization success, see Fig.~\ref{fig:theory}b. Mechanistically, the equilibration timescale is set by the time it takes for the influx of enzyme $J$ across the channel cross-section $w\cdot H$ to reach the enzyme concentration upstream of both channels $C_0$ within the entire channel volume considering a channel length $L$, i.e.~$t_{\mathrm{eq}}=C_0 L w H/J$. Neglecting diffusive influx at $\mathrm{Pe}>1$ considered here, the enzyme influx is given by $J=C_0 U^{\mathrm{eff}}wH$. Therefore, the equilibration time results in
 \begin{equation}\label{eq:t_eq}
t_{\mathrm{eq}}=\frac{L}{U^{\mathrm{eff}}},
\end{equation}
which is in agreement with the numerically determined equilibration timescales, see  Fig.~\ref{fig:theory}b. Note, that the closed expression in Eq.~\eqref{eq:t_eq} hides the fact that the effective enzyme velocity $U^{\mathrm{eff}}$ depends on all three non-dimensional parameters of the diffusion-advection-sorption dynamics, i.e.~the  Péclet number, $\mathrm{Pe}=\bar{U}_{\rm nar}w_{\rm nar}/D$, absorption Damköhler number, $k_a=K_a w_{\rm nar}/D$, and desorption Damköhler number, $k_d=K_d w^2_{\rm nar}/D$. To illustrate their role, we numerically assess homogenization by $\sigma(t)/\sigma(0)$ in two parallel channels for different sets of the three non-dimensional parameters, see Fig.~\ref{fig:theory}c. The channel widths are now fixed to $w_{\mathrm{nar}}=$ \SI{200}{\um} and $w_{\mathrm{wide}}=$ \SI{300}{\um}, respectively, while varying kinetic constants $K_a$, $K_d$ and diffusivity $D$. We find that high absorption Damköhler number, $k_a\gg 1$, and low desorption Damköhler number, $k_d< 10$, increase homogenization. Also, reducing the Péclet number $\mathrm{Pe}$ increases homogenization. The plane with $k_d=4$ is depicted in Fig.~\ref{fig:theory}c, see Movie 6 in Supplementary Material and Supp.~Fig.~13a for the full phase space~\cite{supp}. %
\begin{figure*}[t]
\centering
    \includegraphics[width=0.99\textwidth]{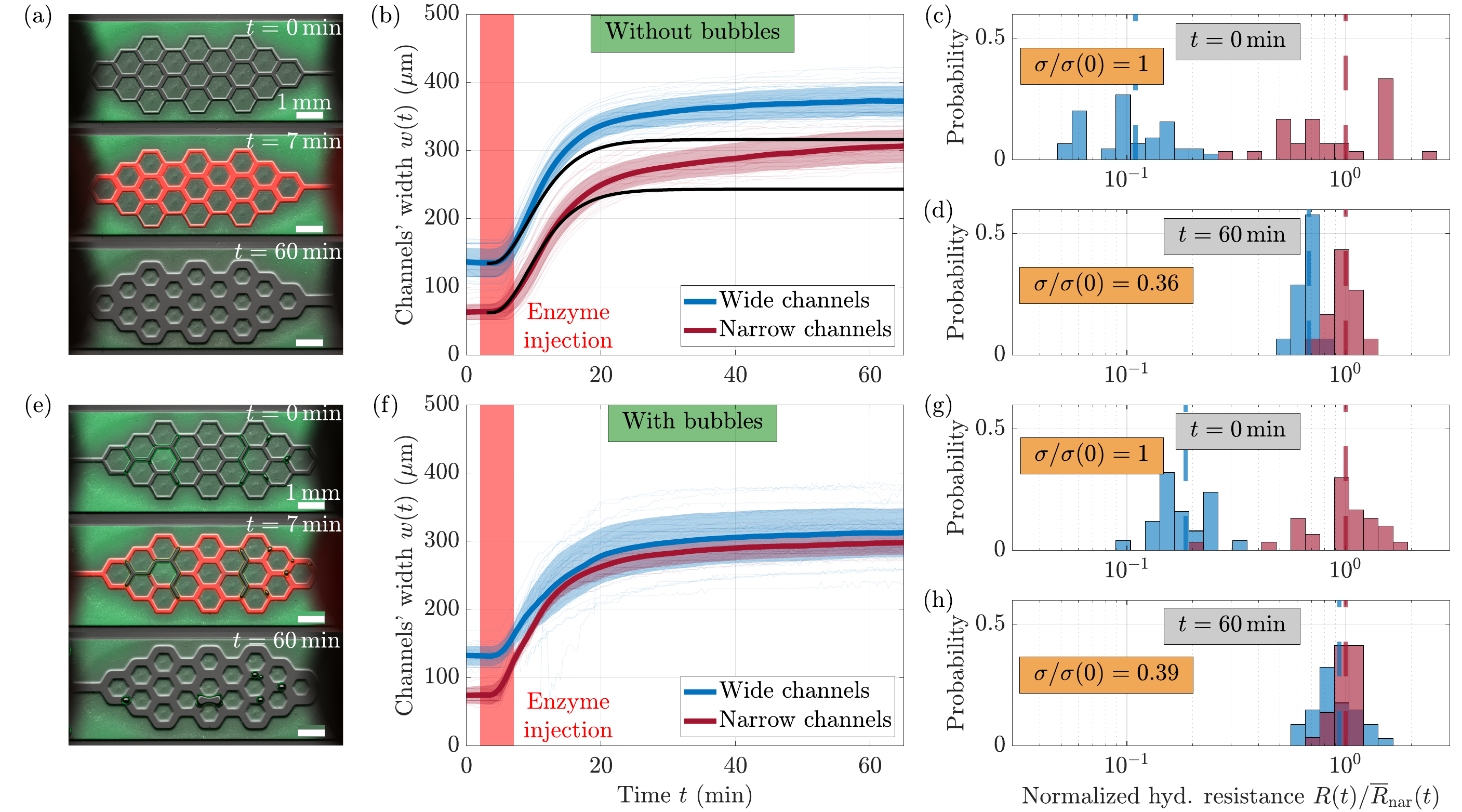} 
\caption{\textbf{Erosion homogenizes complex, looped networks, and works even better with bubbles.} Erosion of a loopy, hexagonal network with an initial bimodal distribution of channel widths perfused without (a,b,c,d) and with (e,f,g,h) air bubbles into wider channels before enzyme pulse (see Movies~3 \& 4 in the Supplemental Material~\cite{supp}). (a,e) Time-lapses of the erosion at $t=0$ (top), $t=\SI{7}{min}$ (middle), and $t=\SI{60}{min}$ (bottom) show the hydrogel (light green) getting eroded by the enzyme (red). With bubbles blocking wide channels, see dark channels, the enzyme does not reach and thus does not erode wide channels hydrogel walls. Scale bar: \SI{1}{\mm}. (b,f) Increase of the channels' width due to the hydrogel erosion. The initially wide (resp.~narrow) channels are depicted in blue (resp.~red). Black lines indicate numerical simulations. (c,d,g,h) Distribution of the channels' hydraulic resistances $R(t)$ normalized by the average hydraulic resistance of the narrow channels $\overline{R}_\mathrm{nar}(t)$, shown at the start of the experiment ($t=0$, c \& g) and after one enzyme pulse ($t=\SI{60}{min}$, d \& h). The value of the normalized standard deviation $\sigma(t)/\sigma(0)$ of the normalized hydraulic resistance $R(t)/\overline{R}_\mathrm{nar}(t)$ of all channels (wide and narrow) is annotated for each distribution.}
\label{fig:3}
\end{figure*}

Again, homogenization is correctly predicted by the time to equilibrate $t_{\rm eq}$ which is inversely proportional to the effective flow velocity, see Supp.~Fig.~13b. Mechanistically, the effective flow velocity $U^{\mathrm{eff}}$ is higher at high absorption kinetics, as high absorption reduces the number of slow enzymes close to the channel wall, thus effectively increasing the flow velocity of the cross-sectionally averaged enzyme concentration. Analogously, low desorption kinetics reduces the inflow of slow enzymes from the wall to the slow streamlines close to the channel wall, thus increasing the effective flow velocity of the cross-sectionally averaged enzyme concentration. We also observe that a reduced Péclet number, i.e.~slower flow, increases homogenization, which is counterintuitive at first sight. However, this can be rationalized by noting that a smaller P\'{e}clet number means a smaller effective diffusivity $D^{\mathrm{eff}}$. Reducing the Péclet number thus reduces the spread of the pulse front due to diffusion in the channels upstream; this, in turn, steepens the incoming pulse front such that, per inflow rate, more enzyme reaches a channel at the onset of the pulse, thereby reducing the time to equilibrate.  
Zooming from the enzyme dynamics within a single channel out to channel networks again, we find that short times to equilibrate to upstream enzyme concentration result in similar absorption of enzyme across channels and thus homogenization by equal erosion.

Yet, it is crucial to note that homogenization is not limited to fast equilibration. For homogenization to take place, in fact, narrow channels only need to erode more than the initial ratio of narrow to wide channel width, $\Delta w_{\text{nar}}/\Delta w_{\text{wide}}>w_{\text{nar}}/w_{\text{wide}}$, according to \Eqref{eq:homo}. Hence, channels may homogenize even if they do not erode equally but only roughly similar, which can be easily reached by adjusting the erosion pulse length. Channel wall erosion only depends on the integral over the square root of the absorbed enzyme, see \Eqref{eq:widthdyn}. Thus, similar erosion of channels within a network is reachable if large times after concentration equilibration dominate wall erosion. Such large times are reached by increasing the pulse duration. We test the transition from heterogenization at low pulse length to homogenization at increased pulse length in parallel channels, where now the parallel channels are partitioned into three section of varying width, see Fig.~\ref{fig:transition}. The very narrow entry section of the narrow channel ensures a much longer equilibration time in the narrow channel compared to the wider one. The small difference in width of the center section between the narrow and the wide channel, i.e.~$w_{\text{nar}}/w_{\text{wide}}\approx 1$, ensures that only a little less erosion in the narrow channel compared to the wide channel already hinders homogenization. Increasing the erosion pulse length relative to the equilibration time indeed promotes homogenization over heterogenization.
Therefore, our theoretical framework predicts robust homogenization as long as the enzyme pulse length $t_{\mathrm{e}}\gg t_{\mathrm{eq}}$ is much longer than the time for enzyme concentration for any individual channel within a network to equilibrate with the enzyme concentration upstream - even independent of the network geometry.

\subsection{Erosion homogenizes complex, looped networks, and works even better with bubbles} 
We put erosion-driven homogenization to the final test by eroding a loopy, hexagonal network of imbalanced channels, with initial widths either drawn from a bimodal distribution centered on $w\simeq\SI{65}{\um}$ and $w\simeq\SI{135}{\um}$ (see Fig.~\ref{fig:3}a and Supp.~Fig.~6a,b~\cite{supp}), or randomly distributed between 50 and \SI{330}{\um}, see Supp.~Fig.~6c~\cite{supp}. We keep all the experimental parameters fixed except for the geometry of the network. Notably, we always set the duration of the enzyme pulse $t_{\mathrm{e}}>t_{\mathrm{eq}}$ to exceed the concentration equilibration time. We observe that all channels erode at the same rate, i.e.~all channels' widths increase at the same rate, see Fig.~\ref{fig:3}b and Movie~3 in the Supplemental Material~\cite{supp}. The histogram of channel resistances shows a clear homogenization after only one enzyme pulse with the normalized standard deviation $\sigma(t)/\sigma(0)$ of normalized resistances shrinking from 1 to 0.36, see Fig.~\ref{fig:3}c,d and Supp.~Fig.~7~\cite{supp}. Numerically solving the solute spread and the erosion dynamics results in excellent agreement with experimental observations in the early time regime, see Fig.~\ref{fig:3}b. Yet, neglecting the enzyme diffusion into the hydrogel in the theoretical descriptions accounts for the deviations at late-time erosion dynamics. Calculating the flow rates using the channel widths before and after erosion shows that the flow rate distribution homogenizes in the experiment shown in Fig.~\ref{fig:3}a-d, with its standard deviation shrinking from 0.31 to 0.22 over a single pulse, see Supp.~Fig.~12a~\cite{supp}, in perfect quantitative agreement with the decline of the standard deviation predicted by numerical simulations of the dynamics, see Supp.~Fig.~12b~\cite{supp}. The simulations reveal that a similar supply to all the channels, regardless of their widths, leads to a similar amount of erosion, which leads to the homogenization of the flow rates, see Supp.~Fig.~12c~\cite{supp}. Altogether, experimental and numerical data on loopy, hexagonal networks confirm the robustness of flow network homogenization via erosion: channels erode at the same rate, driving the homogenization of network-wide resistance and flow rates. 

Although erosion homogenizes channel resistances in complex networks, full equality in resistance is never reached as both wide and narrow channels erode simultaneously. To selectively erode narrow channels only, we inject air bubbles into the network prior to eroding. Instead of carefully plugging the capillary tubing in the channel inlet while the inlet pool is already full of liquid, we plug it while it is mostly empty. This allows the formation of air bubbles, which follow the path of least resistance~\cite{boukellal2009simple}, and thus predominantly reach the low resistance large channels where they get stuck (see Supp.~Fig.~10~\cite{supp} for a comparison of channel resistance and bubble occupation). The subsequently injected enzyme is blocked by the bubbles in the large, low resistance channels, and the enzyme predominantly reaches and erodes narrow channels, see Fig.~\ref{fig:3}e and Movie~4 in the Supplemental Material~\cite{supp}. The erosion of the narrow channels drives them to catch up in width with wider channels, see Fig.~\ref{fig:3}f, and the resistance distribution homogenizes more than without bubbles, see Fig.~\ref{fig:3}g,h. The enhanced homogenization stands out as the initial bimodality of the resistances' distribution disappears in the presence of the bubbles, whereas it is still prominent without, see Fig.~\ref{fig:3}d,h. Once narrow channels reach a width comparable to that of wide channels, bubbles move freely toward their nearest low resistance channel, thus acting as equalizers of resistance by blocking the lowest resistance channels. Once the channel widths exceed $\approx\SI{200}{\um}$, the bubbles are flushed out when the flow rate is \SI{20}
{\ul\per\min}. A lower flow rate allows the bubbles to remain longer in the network, increasing their efficiency to homogenize the channels' widths, see Movie~5 in the Supplemental Material~\cite{supp}. Furthermore, we find that bubbles act robustly as equalizers independent of channel width randomness (see Supp.~Mat.~6~\cite{supp}), as long as the bubbles are large enough to obstruct wide channels and the flow rates small enough not to flush out the bubbles. 

\section{Discussion}
We showed that flowing pulses of an eroding solution into a network with initially imbalanced channel widths induces network self-organization, leading to the homogenization of channel resistances and flows throughout the network. Erosion works robustly as wall erosion only weakly depends, i.e.~by square root, on the exact amount of eroding solute concentration, and thus channels typically erode by a similar amount which leads to the homogenization of resistances simply by the non-linear scaling of the resistance with channel width. The key for similar wall erosion, and thereby homogenization, is the fast equilibration of solute concentration within individual channels compared to pulse length, whose dependence on Péclet and Damköhler numbers we determine analytically. To selectively erode the narrowest channels, we show the injection of bubbles prior to the pulse of the eroding solution as a promising method.

Previous work on adaptive microfluidics used the swelling and contraction of embedded responsive hydrogels to control flow in microchannels by switching channels open or closed~\cite{beebe2000functional,baldi2002hydrogel,liu2002fabrication,richter2003electronically,eddington2004flow,d2018microfluidic,na2022hydrogel}. Yet, no adaptation of network geometry crucial for flow homogenization was attempted. Here, we embed changes in channel geometry and their associated fluid flows due to the progressive but controlled erosion of hydrogel walls, leading to continuous homogenization of the channel resistances, showing that even a simple adaptation, once controlled by timing, can lead to network homogenization. As flow network architecture plays a key role in biological and synthetic active fluids \cite{Jorge.2024dlr,Woodhouse.2016}, the feedback between hydrogel channels and flows introduced here has the potential to broaden the design rules not only for microfluidics but active fluid networks alike.

The homogenization mechanism introduced in this work is based on the local erosion of the channel wall by an eroding chemical transported by the fluid flow. The presence of bubbles usually considered a severe problem to circumvent in fluidic applications~\cite{gravesen1993microfluidics,sung2009prevention,lochovsky2012bubbles}, actually plays to the advantage of homogenization by erosion here. As the hydrogel used in our experiments is biocompatible~\cite{stoberl2023photolithographic}, purposely tailoring the timing of erosive chemical pulsing emerges as a tool for designing adaptive, biomimetic self-controlled fluidic networks for biotechnological applications. In particular the challenge of tissue perfusion in bio-engineered organs could be overcome by the local channel adaptation dynamics introduced here \cite{Stankey.2024,Sexton.2025}.

Nature's tendency is to create disorder. Flow networks in this regard resemble disordered systems in general, and in their statistical properties granular systems in particular \cite{alim2017local}. Counteracting disorder by such simple means as erosion may, therefore, be applicable to other disordered systems.
Note, that, as our theoretical description of the dynamics underlines, flow homogenization by erosion is general and not limited to the specific hydrogel and enzyme used in this work. Neither does erosion homogenize solely in two dimensions, used here for quantitative understanding: the non-linear dependency of channel resistance on channel width prevails in three dimensions, as do the transport dynamics, envisioning applications to three dimensional porous media such as packed-bed reactors. Self-homogenization by pulse-controlled erosion is a generally simple and attractive solution for applications where inhomogeneities cannot be avoided to arise over time. 

\begin{acknowledgments}
This project has received funding from the French Agence Nationale de la Recherche (ANR), under grant ANR-21-CE30-0044, and the Deutsche Forschungsgemeinschaft (DFG, German Research Foundation) under grant AL 1429/5-1 and the INST 95/1634-1 FUGG as well as funding by the IGSSE / TUM Graduate School for IPT SOFT. This research was conducted within the Max Planck School Matter to Life supported by the German Federal Ministry of Education and Research (BMBF) in collaboration with the Max Planck Society.
\end{acknowledgments}

\appendix
\section{Experimental Methods}
The experimental setup consists of microfluidic channels whose side walls are made of a norbornene polyethylene glycol (PEG-NB-8arm) backbone, cross-linked with a peptide chain (KCGPQGIWGQCK-OH, Iris Biotech, Germany). In the presence of the enzyme matrix metalloproteinase-1 (MMP-1, from \textit{Clostridium histolyticum}, Sigma-Aldrich, Germany), the peptide chain is cleaved and the PEG hydrogel erodes. 

Briefly, to fabricate erodible microfluidic channels, \SI{30}{\uL} of an aqueous polymer solution, containing \SI{5}{\mmol\per\liter} PEG-NB-8arm, \SI{20}{\mmol\per\liter} Crosslinker, \SI{3}{\mmol\per\liter} of the photoinitiator lithium phenyl-2,4,6-trimethylbenzoylphosphinate (LAP) and \SI{2}{\mmol\per\liter} of a green fluorescent dye (fluorescein-PEG-Thiol from NanoCS, USA) is flowed into a \SI{400}{\um}-height plastic microfluidic chip ($\mu$-Slide VI 0.4 ibiTreat, Ibidi, Germany). The chip is aligned on a shadow mask with the desired channel geometry and illuminated through a collimated light source at \SI{365}{\nm}; the exposed areas are cross-linked and form the walls of the microfluidic channels~\cite{skinner1997photomask,dietrich2018guiding,stoberl2023photolithographic}. The channels are then washed with PBS (Phosphate Buffer Saline without Calcium nor Magnesium, VWR), and stored at $4^{\circ}{\rm C}$ filled with PBS in humid conditions for at least 5 days. This results in well-defined fluorescent hydrogel walls of \SI{400}{\um} height and up to \SI{10}{\mm} in length, see Fig.~\ref{fig:1}a. 

To erode the hydrogel walls, a solution containing $c_0=\SI{3e-5}{\mol\per\liter}$ MMP-1 and Texas Red-labeled dextran with the same molecular weight, 70\,kDa, is injected at a flow rate of $Q=\SI{20}{\uL\per\min}$ into the microfluidic networks. Beforehand, PBS is flowed into the channels for 30\,min at the same flow rate $Q$ to flush out both the green dye, that had diffused from the hydrogel into the channels during the week-long storage, and bubbles that may have appeared upstream during the tubing connection. Glass syringes of 1 and 10\,mL (CETONI, Germany) are set up on two syringe pumps (Nemesys S, CETONI, Germany), which are controlled by a computer through CETONI Elements software (CETONI, Germany). This software allows for chronological automation of input flows via a script system, enabling perfect repeatability of the injection protocol. Images of the erosion process are taken by fluorescence and bright-field imaging every \SI{20}{\s} with a Hamamatsu ORCA-Flash 4.0 digital camera under an AxioZoom V.16 microscope (Zeiss, Germany), using a Zeiss PlanNeoFluar Z 1x objective. Zeiss Zen 3.2 (blue edition) software was used for imaging. 
\section{Numerical Methods}
To numerically solve for the spread and absorption of a diffusive solute along the channel network, a Crank-Nicolson routine was employed to integrate the dynamics of the cross-sectionally averaged diffusion-advection-sorption dynamics Eqs.~\ref{eq:finaladvdiff}, see~\cite{zhang_transient_2017}, implemented in MATLAB (MathWorks). %
The channel geometry used for the simulation in Figs.~\ref{fig:1}c, d and ~\ref{fig:3}b followed the same design as the hydrogel channel networks: channels were reduced to their skeleton, decomposing every channel into segments which were further decomposed to boxes which is the smallest simulation unit. The connectivity of every channel was encoded into a connectivity matrix, thus defining the geometry of the network. Next, based on the channel width and length, each channel's hydraulic resistance was computed. Then, setting the inflow rate at the inlet and the matching outflow rate at the outlet, Kirchhoff's circuit law was employed, conserving fluid volume at every network node, to compute the flow profile in the channel network. Lastly, the network topology and the flow profile were used to numerically integrate the advection-diffusion-sorption equation along individual channels using the Crank-Nicolson integration routine. All model parameters were directly quantified from experimental data, except the absorption/desorption rate which were matched to fit experiments. At channel junctions, merging and splitting of solute concentration conditions were implemented following~\cite{Meigel.2022}. 
The solute concentration at the inlet is set to vary in time. To this end, we integrated the Taylor dispersion of the solute pulse along the tubing between the upstream switch and the chip inlet. In general, for a rectangular concentration pulse of concentration $\bar{c}_0$ extending from $x_0$ to $x_0-\Delta x$ in one dimension, the concentration profile is given by the result of the spatial superposition i.e.~integral overall solutions for individual $\delta$-peaks,
\begin{eqnarray}\label{entrypulse}
    \bar{c}(x,t)&=&\frac{\bar{c}_0\sqrt{\pi}}{2}\Bigg[\mathrm{erf}{\left(\frac{\left(x-U_{\mathrm{tube}}t-x_0+\Delta x\right)}{\sqrt{4D_{\mathrm{tube}}t}}\right)}\nonumber\\
    &&-\mathrm{erf}{\left(\frac{\left(x-U_{\mathrm{tube}}t-x_0\right)}{\sqrt{4D_{\mathrm{tube}}t}}\right)}\Bigg],
\end{eqnarray}
where, $r_{\mathrm{tube}}$ is the effective tube radius, $U_{\text{tube}}=Q/\pi r_{\text{tube}}^2$ and $D_{\text{tube}}=D \left( 1 + \frac{r_{\text{tube}}^2U_{\text{tube}}^2}{48D^2} \right)$. Both the flow rate $Q$ and the distance between the switch and the chip inlet, $x_0$, are set by the experiments. The tube radius is variable throughout the length of the tube, such that $r_{\text{tube}}$, $D_{\text{tube}}$, and $\bar{c}_0$ are determined by fitting to the data. We fit \Eqref{entrypulse} to the experimentally obtained channel concentration at the channel inlet, normalized by the enzyme concentration injected in experiments, to obtain $\bar{c}_0=0.72$, which is less than one due to the cross-sectional averaging, %
and use these values to calculate the concentration at the channel inlet $x=0$ using \Eqref{entrypulse}. 
At the channel outlet, an open outflow condition was implemented, which estimated transport by advection and diffusion across the last spatial simulation point as in~\cite{Meigel.2022}. The effective coefficients $U^{\rm eff}$ and $D^{\rm eff}$ used for simulations were validated for $\mathrm{Pe}=10$, $k_a=50$ and $k_d=1$ against analytical solutions and plots \cite{zhang_transient_2017} within a single channel.
\bibliography{erosion2}

@article{Jorge.2024dlr, 
year = {2024}, 
title = {{Active hydraulics laws from frustration principles}}, 
author = {Jorge, Camille and Chardac, Amélie and Poncet, Alexis and Bartolo, Denis}, 
journal = {Nature Physics}, 
issn = {1745-2473}, 
doi = {10.1038/s41567-023-02301-2}, 
pages = {303--309}, 
number = {2}, 
volume = {20}
}

@article{Sexton.2025, 
year = {2025}, 
title = {{Rapid model-guided design of organ-scale synthetic vasculature for biomanufacturing}}, 
author = {Sexton, Zachary A. and Rütsche, Dominic and Herrmann, Jessica E. and Hudson, Andrew R. and Sinha, Soham and Du, Jianyi and Shiwarski, Daniel J. and Masaltseva, Anastasiia and Solberg, Fredrik Samdal and Pham, Jonathan and Szafron, Jason M. and Wu, Sean M. and Feinberg, Adam W. and Skylar-Scott, Mark A. and Marsden, Alison L.}, 
journal = {Science (New York, N.Y.)}, 
doi = {10.1126/science.adj6152}, 
pmid = {40504910}, 
pages = {1198--1204}, 
number = {6752}, 
volume = {388}
}

@article{Stankey.2024, 
year = {2024}, 
title = {{Embedding biomimetic vascular networks via coaxial sacrificial writing into functional tissue}}, 
author = {Stankey, Paul P. and Kroll, Katharina T. and Ainscough, Alexander J. and Reynolds, Daniel S. and Elamine, Alexander and Fichtenkort, Ben T. and Uzel, Sebastien G.M. and Lewis, Jennifer A.}, 
journal = {Advanced Materials}, 
issn = {0935-9648}, 
doi = {10.1002/adma.202401528}, 
pmid = {39092638}, 
pages = {e2401528}
}

@article{Woodhouse.2016, 
year = {2016}, 
title = {{Stochastic cycle selection in active flow networks}}, 
author = {Woodhouse, Francis G. and Forrow, Aden and Fawcett, Joanna B. and Dunkel, Jörn}, 
journal = {Proceedings of the National Academy of Sciences}, 
issn = {0027-8424}, 
doi = {10.1073/pnas.1603351113}, 
pmid = {27382186}, 
pmcid = {PMC4961200}, 
eprint = {1607.08015}, 
pages = {8200--8205}, 
number = {29}, 
volume = {113}
}

@Article{Kirkegaard.2020,
  author   = {Kirkegaard, Julius B and Sneppen, Kim},
  journal  = {Phys. Rev. Lett.},
  title    = {{Optimal Transport Flows for Distributed Production Networks}},
  year     = {2020},
  number   = {20},
  pages    = {208101},
  volume   = {124},
  doi      = {10.1103/physrevlett.124.208101},
  fjournal = {Physical Review Letters},
}

@Article{Chang.20175gc,
  author   = {Chang, Shyr-Shea and Tu, Shenyinying and Baek, Kyung In and Pietersen, Andrew and Liu, Yu-Hsiu and Savage, Van M and Hwang, Sheng-Ping L and Hsiai, Tzung K and Roper, Marcus},
  journal  = {PLoS Comput. Biol.},
  title    = {{Optimal occlusion uniformly partitions red blood cells fluxes within a microvascular network}},
  year     = {2017},
  number   = {12},
  pages    = {e1005892},
  volume   = {13},
  doi      = {10.1371/journal.pcbi.1005892},
  fjournal = {PLoS Computational Biology},
}

@Article{beven1982macropores,
  author    = {Beven, Keith and Germann, Peter},
  journal   = {Water Resour. Res.},
  title     = {Macropores and water flow in soils},
  year      = {1982},
  number    = {5},
  pages     = {1311--1325},
  volume    = {18},
  doi       = {10.1029/wr018i005p01311},
  publisher = {Wiley Online Library},
}

@Article{Kramer.2023,
  author   = {Kramer, Felix and Modes, Carl D.},
  journal  = {Phys. Rev. Research},
  title    = {{Biological flow networks: Antagonism between hydrodynamic and metabolic stimuli as driver of topological transitions}},
  year     = {2023},
  number   = {2},
  pages    = {023106},
  volume   = {5},
  doi      = {10.1103/physrevresearch.5.023106},
  fjournal = {Physical Review Research},
}

@Article{Karschau.2020,
  author   = {Karschau, J and Scholich, A and Wise, J and Morales-Naverrete, Hernan and Zerial, Marino and Friedrich, Benjamin M},
  journal  = {PLoS Comput. Biol.},
  title    = {{Resilience of three-dimensional sinusoidal networks in liver tissue}},
  year     = {2020},
  number   = {6},
  pages    = {e1007965},
  volume   = {16},
  doi      = {10.1371/journal.pcbi.1007965},
  fjournal = {PLoS Computational Biology},
}

@Article{Datta.2013,
  author   = {Datta, Sujit S. and Chiang, H. and Ramakrishnan, T. S. and Weitz, David A.},
  journal  = {Phys. Rev. Lett.},
  title    = {{Spatial Fluctuations of Fluid Velocities in Flow through a Three-Dimensional Porous Medium}},
  year     = {2013},
  number   = {6},
  pages    = {064501},
  volume   = {111},
  doi      = {10.1103/physrevlett.111.064501},
  fjournal = {Physical Review Letters},
}

@Article{Matyka.2016,
  author   = {Matyka, Maciej and Gołembiewski, Jarosław and Koza, Zbigniew},
  journal  = {Phys. Rev. E},
  title    = {{Power-exponential velocity distributions in disordered porous media}},
  year     = {2016},
  number   = {1},
  pages    = {013110},
  volume   = {93},
  doi      = {10.1103/physreve.93.013110},
  fjournal = {Physical Review E: Statistical Physics, Plasmas, Fluids, and Related Interdisciplinary Topics},
}

@Article{Lebon.1996,
  author   = {Lebon, L and Oger, L and Leblond, J and Hulin, J P and Martys, N S and Schwartz, L M},
  journal  = {Phys. Fluids},
  title    = {{Pulsed gradient NMR measurements and numerical simulation of flow velocity distribution in sphere packings}},
  year     = {1996},
  issn     = {1070-6631},
  number   = {2},
  pages    = {293--301},
  volume   = {8},
  doi      = {10.1063/1.868839},
  fjournal = {Physics of Fluids},
}

@Article{Kutsovsky.1996,
  author   = {Kutsovsky, Y E and Scriven, L E and Davis, H T and Hammer, B E},
  journal  = {Phys. Fluids},
  title    = {{NMR imaging of velocity profiles and velocity distributions in bead packs}},
  year     = {1996},
  number   = {4},
  pages    = {863--871},
  volume   = {8},
  doi      = {10.1063/1.868867},
  fjournal = {Physics of Fluids},
}

@Article{Shattuck.1997,
  author   = {Shattuck, M. D. and Behringer, R. P. and Johnson, G. A. and Georgiadis, J. G.},
  journal  = {J. Fluid Mech.},
  title    = {{Convection and flow in porous media. Part 1. Visualization by magnetic resonance imaging}},
  year     = {1997},
  pages    = {215--245},
  volume   = {332},
  doi      = {10.1017/s0022112096003990},
  fjournal = {Journal of Fluid Mechanics},
}

@Article{Maier.1998,
  author   = {Maier, R S and Kroll, D M and Kutsovsky, Y E and Davis, H T and Bernard, R S},
  journal  = {Phys. Fluids},
  title    = {{Simulation of flow through bead packs using the lattice Boltzmann method}},
  year     = {1998},
  number   = {1},
  pages    = {60--74},
  volume   = {10},
  doi      = {10.1063/1.869550},
  fjournal = {Physics of Fluids},
}

@Article{Sederman.2001,
  author   = {Sederman, A.J and Gladden, L.F},
  journal  = {Magn. Reson. Imaging},
  title    = {{Magnetic resonance visualisation of single- and two-phase flow in porous media}},
  year     = {2001},
  number   = {3-4},
  pages    = {339--343},
  volume   = {19},
  doi      = {10.1016/s0730-725x(01)00246-6},
  fjournal = {Magnetic Resonance Imaging},
}

@Article{Moroni.2001,
  author   = {Moroni, Monica and Cushman, John H.},
  journal  = {Phys. Fluids},
  title    = {{Statistical mechanics with three-dimensional particle tracking velocimetry experiments in the study of anomalous dispersion. II. Experiments}},
  year     = {2001},
  number   = {1},
  pages    = {81--91},
  volume   = {13},
  doi      = {10.1063/1.1328076},
  fjournal = {Physics of Fluids},
}

@Article{Qi.2021,
  author   = {Qi, Yujia and Roper, Marcus},
  journal  = {Proc. Natl. Acad. Sci. U. S. A.},
  title    = {{Control of low flow regions in the cortical vasculature determines optimal arterio-venous ratios}},
  year     = {2021},
  number   = {34},
  pages    = {e2021840118},
  volume   = {118},
  doi      = {10.1073/pnas.2021840118},
  fjournal = {Proceedings of the National Academy of Sciences},
}

@misc{supp,
note={{See Supplemental Material at [URL will be inserted by publisher] for details on enzyme dynamics, more network architectures self-organizing, flow distributions of self-organizing networks, full parameters sweeps for homogenization, details on the simulation and derivations as well as movies of experimental observations. The experimental data, simulation code, and movies are available in the mediaTUM repository at doi:10.14459/2024mp1752696. The Supplemental Material also includes Ref.~\cite{weber2009effects}}}}

@Article{Kharkar.2013,
  author   = {Kharkar, Prathamesh M. and Kiick, Kristi L. and Kloxin, April M.},
  journal  = {Chem. Soc. Rev.},
  title    = {{Designing degradable hydrogels for orthogonal control of cell microenvironments}},
  year     = {2013},
  number   = {17},
  pages    = {7335--7372},
  volume   = {42},
  doi      = {10.1039/c3cs60040h},
  fjournal = {Chemical Society Reviews},
}

@Article{Murray.1926krhf,
  author   = {Murray, Cecil D},
  journal  = {Proc. Natl. Acad. Sci. U. S. A.},
  title    = {{The Physiological Principle of Minimum Work: I. The Vascular System and the Cost of Blood Volume}},
  year     = {1926},
  issn     = {0027-8424},
  number   = {3},
  pages    = {207--214},
  volume   = {12},
  doi      = {10.1073/pnas.12.3.207},
  fjournal = {Proceedings of the National Academy of Sciences},
}

@Article{Liese.2021,
  author   = {Liese, Susanne and Mahadevan, L. and Carlson, Andreas},
  journal  = {Europhys. Lett.},
  title    = {{Balancing efficiency and homogeneity of biomaterial transport in networks}},
  year     = {2021},
  number   = {5},
  pages    = {58001},
  volume   = {135},
  doi      = {10.1209/0295-5075/135/58001},
  fjournal = {Europhysics Letters},
}

@Article{boddy2009saprotrophic,
  author    = {Boddy, Lynne and Hynes, Juliet and Bebber, Daniel P and Fricker, Mark D},
  journal   = {Mycoscience},
  title     = {Saprotrophic cord systems: dispersal mechanisms in space and time},
  year      = {2009},
  number    = {1},
  pages     = {9--19},
  volume    = {50},
  doi       = {10.1007/s10267-008-0450-4},
  publisher = {Elsevier},
}

@Article{sun2019hierarchical,
  author    = {Sun, Hongtao and Zhu, Jian and Baumann, Daniel and Peng, Lele and Xu, Yuxi and Shakir, Imran and Huang, Yu and Duan, Xiangfeng},
  journal   = {Nat. Rev. Mater.},
  title     = {Hierarchical 3D electrodes for electrochemical energy storage},
  year      = {2019},
  number    = {1},
  pages     = {45--60},
  volume    = {4},
  doi       = {https://doi.org/10.1038/s41578-018-0069-9},
  fjournal  = {Nature Reviews Materials},
  publisher = {Nature Publishing Group UK London},
}

@Article{bowen1995theoretical,
  author    = {Bowen, W Richard and Jenner, Frank},
  journal   = {Adv. Colloid Interfac.},
  title     = {Theoretical descriptions of membrane filtration of colloids and fine particles: an assessment and review},
  year      = {1995},
  pages     = {141--200},
  volume    = {56},
  doi       = {10.1016/0001-8686(94)00232-2},
  fjournal  = {Advances in Colloid and Interface Science},
  publisher = {Elsevier},
}

@Article{tien1979advances,
  author    = {Tien, Chi and Payatakes, Alkiviades C},
  journal   = {AIChE J.},
  title     = {Advances in deep bed filtration},
  year      = {1979},
  number    = {5},
  pages     = {737--759},
  volume    = {25},
  doi       = {10.1002/aic.690250502},
  fjournal  = {AIChE Journal},
  publisher = {Wiley Online Library},
}

@Article{fenech2019microfluidic,
  author    = {Fenech, Marianne and Girod, Vincent and Claveria, Viviana and Meance, Sebastien and Abkarian, Manouk and Charlot, Benoit},
  journal   = {Lab Chip},
  title     = {Microfluidic blood vasculature replicas using backside lithography},
  year      = {2019},
  number    = {12},
  pages     = {2096--2106},
  volume    = {19},
  doi       = {10.1039/c9lc00254e},
  fjournal  = {Lab on a Chip},
  publisher = {Royal Society of Chemistry},
}

@Article{wong2012microfluidic,
  author    = {Wong, Keith HK and Chan, Juliana M and Kamm, Roger D and Tien, Joe},
  journal   = {Annu. Rev. Biomed. Eng.},
  title     = {Microfluidic models of vascular functions},
  year      = {2012},
  number    = {1},
  pages     = {205--230},
  volume    = {14},
  doi       = {10.1146/annurev-bioeng-071811-150052},
  fjournal  = {Annual Review of Biomedical Engineering},
  publisher = {Annual Reviews},
}

@Article{sebastian2018microfluidics,
  author    = {Sebastian, Bernhard and Dittrich, Petra S},
  journal   = {Annu. Rev. Fluid Mech.},
  title     = {Microfluidics to mimic blood flow in health and disease},
  year      = {2018},
  number    = {1},
  pages     = {483--504},
  volume    = {50},
  doi       = {10.1146/annurev-fluid-010816-060246},
  fjournal  = {Annual Review of Fluid Mechanics},
  publisher = {Annual Reviews},
}

@Article{tero2010rules,
  author    = {Tero, Atsushi and Takagi, Seiji and Saigusa, Tetsu and Ito, Kentaro and Bebber, Dan P and Fricker, Mark D and Yumiki, Kenji and Kobayashi, Ryo and Nakagaki, Toshiyuki},
  journal   = {Science},
  title     = {Rules for biologically inspired adaptive network design},
  year      = {2010},
  number    = {5964},
  pages     = {439--442},
  volume    = {327},
  doi       = {10.1126/science.1177894},
  publisher = {American Association for the Advancement of Science},
}

@Article{kramar2021encoding,
  author    = {Kramar, Mirna and Alim, Karen},
  journal   = {Proc. Natl. Acad. Sci. U. S. A.},
  title     = {Encoding memory in tube diameter hierarchy of living flow network},
  year      = {2021},
  number    = {10},
  pages     = {e2007815118},
  volume    = {118},
  doi       = {10.1073/pnas.2007815118},
  fjournal  = {Proceedings of the National Academy of Sciences},
  publisher = {National Acad Sciences},
}

@Article{Taylor.1953,
  author   = {Taylor, Geoffrey Ingram},
  journal  = {Proc. R. Soc. London, A},
  title    = {{Dispersion of soluble matter in solvent flowing slowly through a tube}},
  year     = {1953},
  issn     = {0080-4630},
  number   = {1137},
  pages    = {186--203},
  volume   = {219},
  doi      = {10.1098/rspa.1953.0139},
}

@Article{Aris.1956,
  author   = {Aris, R.},
  journal  = {Proc. R. Soc. London, A},
  title    = {{On the dispersion of a solute in a fluid flowing through a tube}},
  year     = {1956},
  issn     = {0080-4630},
  number   = {1200},
  pages    = {67--77},
  volume   = {235},
  doi      = {10.1098/rspa.1956.0065},
}

@Article{Meigel.2022,
  author   = {Meigel, Felix J. and Darwent, Thomas and Bastin, Leonie and Goehring, Lucas and Alim, Karen},
  journal  = {Nat. Commun.},
  title    = {{Dispersive transport dynamics in porous media emerge from local correlations}},
  year     = {2022},
  number   = {1},
  pages    = {5885},
  volume   = {13},
  doi      = {10.1038/s41467-022-33485-5},
  fjournal = {Nature Communications},
}

@Article{Bruus.2011,
  author   = {Bruus, Henrik},
  journal  = {Lab Chip},
  title    = {{Acoustofluidics 1: Governing equations in microfluidics}},
  year     = {2011},
  issn     = {1473-0197},
  number   = {22},
  pages    = {3742--3751},
  volume   = {11},
  doi      = {10.1039/c1lc20658c},
  fjournal = {Lab on a Chip},
  pmid     = {22011885},
}

@Article{boukellal2009simple,
  author    = {Boukellal, Hakim and Selimovi{\'c}, {\v{S}}eila and Jia, Yanwei and Cristobal, Galder and Fraden, Seth},
  journal   = {Lab Chip},
  title     = {Simple, robust storage of drops and fluids in a microfluidic device},
  year      = {2009},
  number    = {2},
  pages     = {331--338},
  volume    = {9},
  doi       = {10.1039/b808579j},
  fjournal  = {Lab on a Chip},
  publisher = {Royal Society of Chemistry},
}

@Article{marbach2023vein,
  author    = {Marbach, Sophie and Ziethen, Noah and Bastin, Leonie and B{\"a}uerle, Felix K and Alim, Karen},
  journal   = {Elife},
  title     = {Vein fate determined by flow-based but time-delayed integration of network architecture},
  year      = {2023},
  pages     = {e78100},
  volume    = {12},
  doi       = {10.7554/elife.78100},
  publisher = {eLife Sciences Publications Limited},
}

@Article{isogai2001vascular,
  author    = {Isogai, Sumio and Horiguchi, Masaharu and Weinstein, Brant M},
  journal   = {Dev. Biol.},
  title     = {The vascular anatomy of the developing zebrafish: an atlas of embryonic and early larval development},
  year      = {2001},
  number    = {2},
  pages     = {278--301},
  volume    = {230},
  doi       = {10.1006/dbio.2000.9995},
  fjournal  = {Developmental Biology},
  publisher = {Elsevier},
}

@Article{lutolf2003synthetic,
  author    = {Lutolf, Matthias P and Lauer-Fields, Janelle L and Schmoekel, Hugo G and Metters, Andrew T and Weber, Franz E and Fields, Greg B and Hubbell, Jeffrey A},
  journal   = {Proc. Natl. Acad. Sci. U. S. A.},
  title     = {Synthetic matrix metalloproteinase-sensitive hydrogels for the conduction of tissue regeneration: engineering cell-invasion characteristics},
  year      = {2003},
  number    = {9},
  pages     = {5413--5418},
  volume    = {100},
  doi       = {10.1073/pnas.0737381100},
  fjournal  = {Proceedings of the National Academy of Sciences},
  publisher = {National Acad Sciences},
}

@Article{gaspers1994enzymes,
  author    = {Gaspers, Pamela B and Robertson, Channing R and Gast, Alice P},
  journal   = {Langmuir},
  title     = {Enzymes on immobilized substrate surfaces: diffusion},
  year      = {1994},
  number    = {8},
  pages     = {2699--2704},
  volume    = {10},
  doi       = {10.1021/la00020a032},
  publisher = {ACS Publications},
}

@Article{patterson2010enhanced,
  author    = {Patterson, Jennifer and Hubbell, Jeffrey Alan},
  journal   = {Biomaterials},
  title     = {Enhanced proteolytic degradation of molecularly engineered PEG hydrogels in response to MMP-1 and MMP-2},
  year      = {2010},
  number    = {30},
  pages     = {7836--7845},
  volume    = {31},
  doi       = {10.1016/j.biomaterials.2010.06.061},
  publisher = {Elsevier},
}

@Article{nagase1996human,
  author    = {Nagase, Hideaki and Fields, Gregg B},
  journal   = {Pept. Sci.},
  title     = {Human matrix metalloproteinase specificity studies using collagen sequence-based synthetic peptides},
  year      = {1996},
  number    = {4},
  pages     = {399--416},
  volume    = {40},
  doi       = {10.1002/(sici)1097-0282(1996)40:4<399::aid-bip5>3.3.co;2-7},
  fjournal  = {Peptide Science},
  publisher = {Wiley Online Library},
}

@Article{dietrich2018guiding,
  author    = {Dietrich, Miriam and Le Roy, Hugo and Br{\"u}ckner, David B and Engelke, Hanna and Zantl, Roman and R{\"a}dler, Joachim O and Broedersz, Chase P},
  journal   = {Soft Matter},
  title     = {Guiding 3D cell migration in deformed synthetic hydrogel microstructures},
  year      = {2018},
  number    = {15},
  pages     = {2816--2826},
  volume    = {14},
  doi       = {10.1039/c8sm00018b},
  publisher = {The Royal Society of Chemistry},
}

@InBook{skinner1997photomask,
  author    = {Skinner, John G. and Groves, Timothy R. and Novembre, Anthony and Pfeiffer, Hans and Singh, Rajeev},
  pages     = {377--474},
  publisher = {SPIE PRESS},
  title     = {Photomask Fabrication Procedures and Limitations},
  year      = {1997},
  isbn      = {9781510607965},
  volume    = {1},
  booktitle = {Handbook of Microlithography, Micromachining, and Microfabrication. Volume 1: Microlithography},
  doi       = {https://doi.org/10.1117/3.2265070.ch5},
  journal   = {Handbook of Microlithography, Micromachining, and Microfabrication, SPIE Optical Engineering Press, Bellingham, WA},
}

@Article{alim2017mechanism,
  author    = {Alim, Karen and Andrew, Natalie and Pringle, Anne and Brenner, Michael P},
  journal   = {Proc. Natl. Acad. Sci. U. S. A.},
  title     = {Mechanism of signal propagation in Physarum polycephalum},
  year      = {2017},
  number    = {20},
  pages     = {5136--5141},
  volume    = {114},
  doi       = {https://doi.org/10.1073/pnas.1618114114},
  fjournal  = {Proceedings of the National Academy of Sciences},
  publisher = {National Acad Sciences},
}

@Article{alim2017local,
  author    = {Alim, Karen and Parsa, Shima and Weitz, David A and Brenner, Michael P},
  journal   = {Phys. Rev. Lett.},
  title     = {Local pore size correlations determine flow distributions in porous media},
  year      = {2017},
  number    = {14},
  pages     = {144501},
  volume    = {119},
  doi       = {10.1103/physrevlett.119.144501},
  fjournal  = {Physical Review Letters},
  publisher = {APS},
}

@Article{meigel2018flow,
  author    = {Meigel, Felix J and Alim, Karen},
  journal   = {J. Roy. Soc. . Interface},
  title     = {Flow rate of transport network controls uniform metabolite supply to tissue},
  year      = {2018},
  number    = {142},
  pages     = {20180075},
  volume    = {15},
  doi       = {10.1098/rsif.2018.0075},
  fjournal  = {Journal of The Royal Society Interface},
  publisher = {The Royal Society},
}

@Article{gavrilchenko2021distribution,
  author    = {Gavrilchenko, Tatyana and Katifori, Eleni},
  journal   = {Phys. Rev. Lett.},
  title     = {Distribution networks achieve uniform perfusion through geometric self-organization},
  year      = {2021},
  number    = {7},
  pages     = {078101},
  volume    = {127},
  doi       = {10.1103/physrevlett.127.078101},
  fjournal  = {Physical Review Letters},
  publisher = {APS},
}

@Article{simonin2012physiological,
  author    = {Simonin, Anna and Palma-Guerrero, Javier and Fricker, Mark and Glass, N Louise},
  journal   = {Eukaryot. Cell},
  title     = {Physiological significance of network organization in fungi},
  year      = {2012},
  number    = {11},
  pages     = {1345--1352},
  volume    = {11},
  doi       = {10.1128/ec.00213-12},
  fjournal  = {Eukaryotic Cell},
  publisher = {Am Soc Microbiol},
}

@Article{chang2019microvascular,
  author    = {Chang, Shyr-Shea and Roper, Marcus},
  journal   = {J. Theor. Biol.},
  title     = {Microvascular networks with uniform flow},
  year      = {2019},
  pages     = {48--64},
  volume    = {462},
  doi       = {10.1016/j.jtbi.2018.10.049},
  fjournal  = {Journal of Theoretical Biology},
  publisher = {Elsevier},
}

@Article{beebe2000functional,
  author    = {Beebe, David J and Moore, Jeffrey S and Bauer, Joseph M and Yu, Qing and Liu, Robin H and Devadoss, Chelladurai and Jo, Byung-Ho},
  journal   = {Nature},
  title     = {Functional hydrogel structures for autonomous flow control inside microfluidic channels},
  year      = {2000},
  number    = {6778},
  pages     = {588--590},
  volume    = {404},
  doi       = {10.1038/35007047},
  publisher = {Nature Publishing Group UK London},
}

@Article{eddington2004flow,
  author    = {Eddington, David T and Beebe, David J},
  journal   = {Adv. Drug Deliver. Rev.},
  title     = {Flow control with hydrogels},
  year      = {2004},
  number    = {2},
  pages     = {199--210},
  volume    = {56},
  doi       = {10.1016/j.addr.2003.08.013},
  fjournal  = {Advanced Drug Delivery Reviews},
  publisher = {Elsevier},
}

@Article{weber2009effects,
  author    = {Weber, Laney M. and Lopez, Christina G. and Anseth, Kristi S.},
  journal   = {J. Biomed. Mater. Res. A},
  title     = {Effects of PEG hydrogel crosslinking density on protein diffusion and encapsulated islet survival and function},
  year      = {2009},
  issn      = {1552-4965},
  month     = jun,
  number    = {3},
  pages     = {720--729},
  volume    = {90},
  doi       = {10.1002/jbm.a.32134},
  publisher = {Wiley Online Library},
}

@InProceedings{baldi2002hydrogel,
  author       = {Baldi, A. and Yuandong Gu and Loftness, P.E. and Siegel, R.A. and Ziaie, B.},
  booktitle    = {Technical Digest. MEMS 2002 IEEE International Conference. Fifteenth IEEE International Conference on Micro Electro Mechanical Systems (Cat. No.02CH37266)},
  title        = {A hydrogel-actuated smart microvalve with a porous diffusion barrier back-plate for active flow control},
  year         = {2002},
  organization = {IEEE},
  pages        = {105--108},
  publisher    = {IEEE},
  series       = {MEMSYS-02},
  collection   = {MEMSYS-02},
  doi          = {10.1109/MEMSYS.2002.984101},
}

@Article{d2018microfluidic,
  author    = {D'eramo, Lo{\"\i}c and Chollet, Benjamin and Leman, Marie and Martwong, Ekkachai and Li, Mengxing and Geisler, Hubert and Dupire, Jules and Kerdraon, Margaux and Vergne, Cl{\'e}mence and Monti, Fabrice and others},
  journal   = {Microsyst. \& Nanoeng.},
  title     = {Microfluidic actuators based on temperature-responsive hydrogels},
  year      = {2018},
  number    = {1},
  pages     = {1--7},
  volume    = {4},
  doi       = {10.1038/micronano.2017.69},
  publisher = {Nature Publishing Group},
}

@Article{richter2003electronically,
  author    = {Richter, Andreas and Kuckling, Dirk and Howitz, Steffen and Gehring, Thomas and Arndt, K-F},
  journal   = {J. Microelectromech. S.},
  title     = {Electronically controllable microvalves based on smart hydrogels: magnitudes and potential applications},
  year      = {2003},
  number    = {5},
  pages     = {748--753},
  volume    = {12},
  doi       = {10.1109/jmems.2003.817898},
  fjournal  = {Journal of Microelectromechanical Systems},
  publisher = {IEEE},
}

@Article{goy2019microfluidics,
  author    = {Goy, Carla B and Chaile, Roberto E and Madrid, Rossana E},
  journal   = {React. Funct. Polym.},
  title     = {Microfluidics and hydrogel: A powerful combination},
  year      = {2019},
  pages     = {104314},
  volume    = {145},
  doi       = {10.1016/j.reactfunctpolym.2019.104314},
  fjournal  = {Reactive and Functional Polymers},
  publisher = {Elsevier},
}

@Article{paratore2022reconfigurable,
  author    = {Paratore, Federico and Bacheva, Vesna and Bercovici, Moran and Kaigala, Govind V},
  journal   = {Nat. Rev. Chem.},
  title     = {Reconfigurable microfluidics},
  year      = {2022},
  number    = {1},
  pages     = {70--80},
  volume    = {6},
  doi       = {https://doi.org/10.1038/s41570-021-00343-9},
  fjournal  = {Nature Reviews Chemistry},
  publisher = {Nature Publishing Group UK London},
}

@Article{na2022hydrogel,
  author    = {Na, Hyeonuk and Kang, Yong-Woo and Park, Chang Seo and Jung, Sohyun and Kim, Ho-Young and Sun, Jeong-Yun},
  journal   = {Science},
  title     = {Hydrogel-based strong and fast actuators by electroosmotic turgor pressure},
  year      = {2022},
  number    = {6590},
  pages     = {301--307},
  volume    = {376},
  doi       = {10.1126/science.abm7862},
  publisher = {American Association for the Advancement of Science},
}

@Article{liu2002fabrication,
  author    = {Liu, Robin H and Yu, Qing and Beebe, David J},
  journal   = {J. Microelectromech. S.},
  title     = {Fabrication and characterization of hydrogel-based microvalves},
  year      = {2002},
  number    = {1},
  pages     = {45--53},
  volume    = {11},
  doi       = {10.1109/84.982862},
  fjournal  = {Journal of Microelectromechanical Systems},
  publisher = {IEEE},
}

@Article{zhang_transient_2017,
  author    = {Zhang, Li and Hesse, Marc A. and Wang, Moran},
  journal   = {J. Fluid Mech.},
  title     = {Transient solute transport with sorption in {Poiseuille} flow},
  year      = {2017},
  issn      = {0022-1120, 1469-7645},
  month     = sep,
  note      = {arXiv:1909.08720 [physics]},
  pages     = {295--332},
  volume    = {828},
  doi       = {https://doi.org/10.1017/jfm.2017.546},
  fjournal  = {Journal of Fluid Mechanics},
  publisher = {Cambridge University Press (CUP)},
  url       = {http://arxiv.org/abs/1909.08720},
  urldate   = {2023-03-06},
}

@Article{zareei_temporal_2022,
  author     = {Zareei, Ahmad and Pan, Deng and Amir, Ariel},
  journal    = {Phys. Rev. Lett.},
  title      = {Temporal {Evolution} of {Erosion} in {Pore} {Networks}: {From} {Homogenization} to {Instability}},
  year       = {2022},
  issn       = {0031-9007, 1079-7114},
  month      = jun,
  number     = {23},
  pages      = {234501},
  volume     = {128},
  doi        = {10.1103/PhysRevLett.128.234501},
  fjournal   = {Physical Review Letters},
  shorttitle = {Temporal {Evolution} of {Erosion} in {Pore} {Networks}},
  url        = {https://link.aps.org/doi/10.1103/PhysRevLett.128.234501},
  urldate    = {2023-03-06},
}

@Article{stoberl2023photolithographic,
  author    = {St{\"o}berl, Stefan and Balles, Miriam and Kellerer, Thomas and R{\"a}dler, Joachim O},
  journal   = {Lab Chip},
  title     = {Photolithographic microfabrication of hydrogel clefts for cell invasion studies},
  year      = {2023},
  number    = {7},
  pages     = {1886--1895},
  volume    = {23},
  doi       = {10.1039/d2lc01105k},
  fjournal  = {Lab on a Chip},
  publisher = {Royal Society of Chemistry},
}

@Article{sung2009prevention,
  author    = {Sung, Jong Hwan and Shuler, Michael L},
  journal   = {Biomed. Microdevices},
  title     = {Prevention of air bubble formation in a microfluidic perfusion cell culture system using a microscale bubble trap},
  year      = {2009},
  pages     = {731--738},
  volume    = {11},
  doi       = {10.1007/s10544-009-9286-8},
  publisher = {Springer},
}

@Article{gravesen1993microfluidics,
  author    = {Gravesen, Peter and Branebjerg, Jens and Jensen, O S{\o}nderg{\aa}rd},
  journal   = {J. Micromech. Microeng.},
  title     = {Microfluidics-a review},
  year      = {1993},
  number    = {4},
  pages     = {168},
  volume    = {3},
  doi       = {10.1088/0960-1317/3/4/002},
  publisher = {IOP Publishing},
}

@Article{lochovsky2012bubbles,
  author    = {Lochovsky, Conrad and Yasotharan, Sanjesh and G{\"u}nther, Axel},
  journal   = {Lab Chip},
  title     = {Bubbles no more: in-plane trapping and removal of bubbles in microfluidic devices},
  year      = {2012},
  number    = {3},
  pages     = {595--601},
  volume    = {12},
  doi       = {10.1039/c1lc20817a},
  fjournal  = {Lab on a Chip},
  publisher = {Royal Society of Chemistry},
}

\end{document}


\maketitle

\thispagestyle{empty}
\beginsupplement
\section{Diffusion coefficient of the enzyme in the hydrogel}
\label{sec:sup_diffusion}
To assess the diffusion coefficient of the enzyme within the hydrogel from our experimental data, we solve the 1D diffusion equation with a constant source of enzyme of concentration $c_0$ at the wall-channel interface, here denoted by $y_0$. We then obtain the concentration profile
\begin{equation}
    c(y,t) = c_0 \left(1-\mathrm{erf}\left(\dfrac{y-y_0}{\sqrt{4Dt}}\right)\right),
\end{equation}
with $D$ the diffusion coefficient of the enzyme in the hydrogel. 

\begin{figure}[ht]
\begin{center}
\includegraphics[trim = 0mm 0mm 0mm 0mm, clip, width=1.0\textwidth, angle=0]{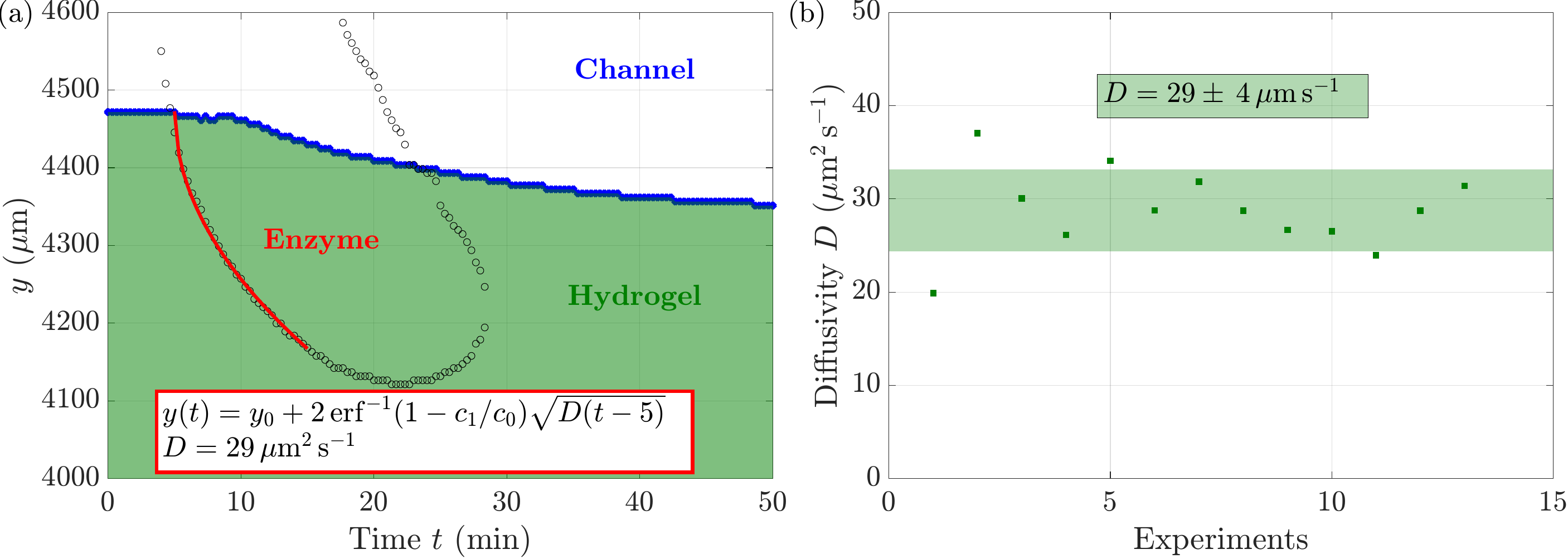}
\caption{\textbf{Measurement of the enzyme diffusion coefficient inside the hydrogel.} (a) Diffusion of the enzyme inside the hydrogel: the diffusive profile can be accurately fitted by Eq.~\eqref{eq:diff}. Data obtained with a double-channel network. (b) Values of the diffusion coefficient measured in different experiments. We obtain $D=\SI{29\pm4}{\square\um\per\s}$ for the diffusivity in the hydrogel.}
\label{fig:diffusion}
\end{center}
\end{figure}

The experimentally obtained enzyme profiles within the hydrogel [black circles in Fig.~\ref{fig:diffusion}(a) and Main~Fig.~2(a)] are isoconcentration lines of enzyme at $c_1=\SI{1.5e-6}{\mol\per\liter}$. On those lines, we thus have
\begin{equation}
    \mathrm{erf}\left(\dfrac{y-y_0}{\sqrt{4Dt}}\right) = 1 - \dfrac{c_1}{c_0},
\end{equation}
from which we can deduce
\begin{equation}
\label{eq:diff}
    y(t) = y_0 + 2~\mathrm{erf}^{-1}\left(1 - \dfrac{c_1}{c_0}\right) \sqrt{D} \sqrt{t}.
\end{equation}
The concentration $c_0$ of the source is $c_0\simeq\SI{1.5e-5}{\mol\per\liter}$.
Thus, fitting the diffusive profiles by a square root $y(t)\propto \sqrt{t}$ yields the value of the diffusivity. The square root fits are in excellent agreement with the data, see Fig.~\ref{fig:diffusion}(a). Repeating the fit over multiple experiments, we obtain $D=\SI{29\pm4}{\square\um\per\s}$ (mean $\pm$ std. deviation), consistent with the literature: similar diffusivities have been measured for the diffusion of similarly sized proteins in PEG hydrogels~\cite{weber2009effects}, and slightly higher diffusivities have been found for a similar collagenase enzyme in liquid buffers~\cite{gaspers1994enzymes}. Individual diffusivity values for different experiments are plotted in Fig.~\ref{fig:diffusion}(b). %
\section{Robust scaling of hydrogel erosion rate in different experimental conditions}
\begin{figure}[ht]
\begin{center}
\includegraphics[width=1.0\textwidth]{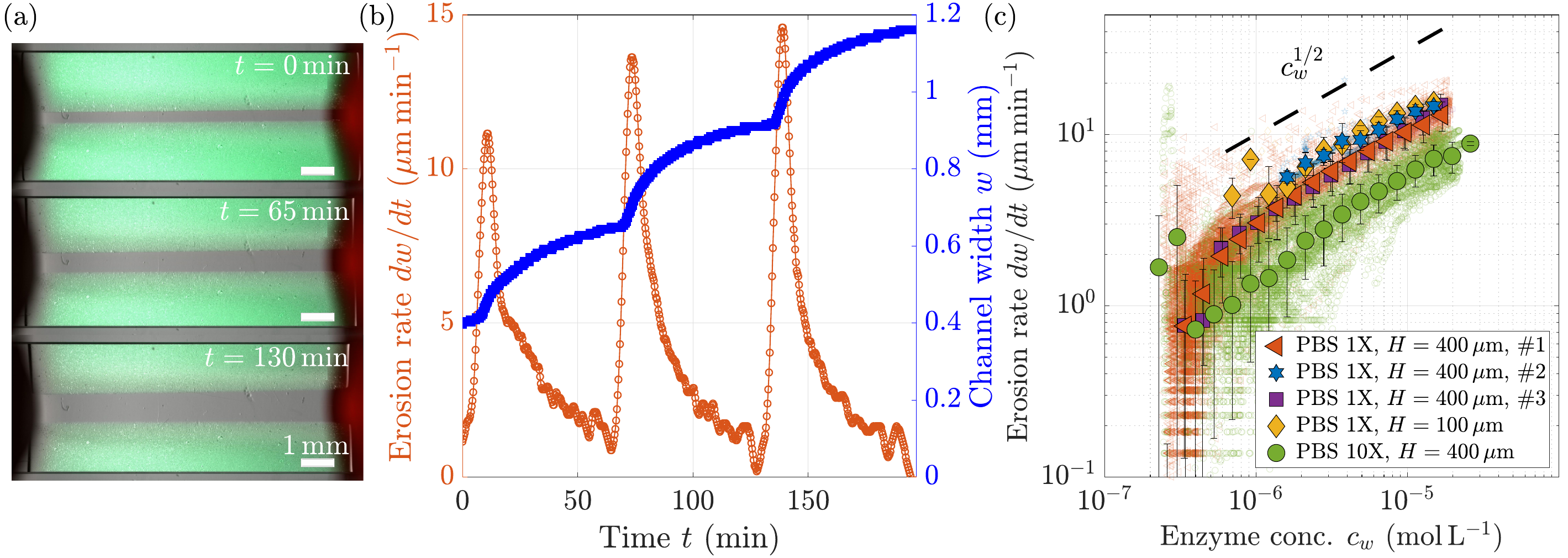}
\caption{\textbf{Erosion rate of the hydrogel in different conditions.} (a) Time-lapse of the erosion of a single channel network of PEG-NB (polyethylene glycol norbornene), see Movie~1 in the Supplemental Material. The channel has a height $H=\SI{400}{\um}$, a length $L\simeq\SI{1}{\mm}$ and an initial width $w\simeq\SI{400}{\um}$. (b) Time-evolution of the erosion rate $dw/dt$ (in orange) for the single-channel network, as three pulses of enzyme are flowed into the device. The channel width $w$, displayed in blue for reference, is the same as in Main~Fig.~2(b). (c) Erosion rate $dw/dt$ of the hydrogel as a function of the local enzyme concentration at the wall $c_w$, in different conditions. All experiments show a square-root dependency on the enzyme concentration. A higher PBS concentration in the hydrogel decreases the erosion speed of the hydrogel, whereas the hydrogel height $H$ is not found to have any influence.}
\label{fig:erosion_rate}
\end{center}
\end{figure}

We find the erosion rate scaling $dw/dt\propto\sqrt{c_w}$, see Main~Eq.~(2), to be valid with a smaller hydrogel height $H$ [yellow diamonds in Fig.~\ref{fig:erosion_rate}(c)], and for devices made using a higher concentration of PBS in the initial PEG-NB solution [green circles in Fig.~\ref{fig:erosion_rate}(c)]. We find $\chi_0=\SI{3.7\pm1e3}{\um\per\min\mol\tothe{-1/2}\liter\tothe{1/2}}$ as a proportional factor for the PBS 1X, whether the height is 100 or \SI{400}{\um}, and $\chi_0=\SI{1.8\pm0.4e3}{\um\per\min\mol\tothe{-1/2}\liter\tothe{1/2}}$ for the PBS 10X. As expected, the hydrogel height does not have any influence on the erosion rate, but the increased buffer concentration, meaning a higher concentration of salts in the hydrogel, is found to have an effect: the erosion is two to three times slower in the latter case. Qualitatively, the hydrogel containing a higher concentration of salts, i.e.~prepared with PBS 10X, is stiffer than the one prepared with a regular concentration of PBS buffer. This implies that the mesh size of this hydrogel is smaller, lowering the diffusion coefficient and the erosion rate. %

\section{Kinetic constants of MMP-1 enzyme}
\label{sec:sup_kinetic_constants}

From the Michaelis-Menten erosion dynamics, we derive the kinetic constant to be
\begin{equation}
k_\mathrm{cat}=\dfrac{\chi_0^2 A_0}{D}.
\end{equation}
As we measure $D\simeq\SI{29\pm4}{\square\um\per\s}$ in the hydrogel from the kymographs (see Supp.~Mat.~\ref{sec:sup_diffusion}), and  $\chi_0=\SIrange{1.4e3}{4.7e3}{\um\per\min\per\mol\tothe{1/2}\liter\tothe{1/2}}$ considering all the data sets in Fig.~\ref{fig:erosion_rate}(c), we can deduce the kinetic constant $k_\mathrm{cat}=\SIrange{0.4}{4.2}{\per\s}$. The comparison to the literature done in Tab.~\ref{tab:kcat}, with close but different substrates, highlights a small discrepancy of less than one order of magnitude, which translates to an even smaller difference on the enzyme concentration. One cannot expect more precision from a simple scaling law analysis of the problem. %

\begin{table}[t]
\centering
\begin{tabularx}{0.96\textwidth}{c c c}
  \toprule
  Substrate & Kinetic constant $k_\mathrm{cat}$  (\si{\per\s})& Reference\\
  \midrule
    Ac-GCRD-GPQG IWGQ-DRCG-NH2 & \num{0.51\pm0.10} & Lutolf \textit{et al.}~\cite{lutolf2003synthetic}\\
    Ac-GCRE-GPQG IWGQ-ERCG-NH2 & \num{0.65\pm0.13} & Patterson and Hubell~\cite{patterson2010enhanced}\\
    GPQG IWGQ & \num{0.36} & Nagase and Fields~\cite{nagase1996human}\\
  PEG750 + Ac-GCRD-GPQG IWGQ-DRCG-NH2 & \num{0.64\pm0.13} & Lutolf \textit{et al.}~\cite{lutolf2003synthetic}\\
    PEG5000 + Ac-GCRD-GPQG IWGQ-DRCG-NH2 & \num{0.83\pm0.17} & Lutolf \textit{et al.}~\cite{lutolf2003synthetic}\\
    (8arm)PEG20k + KCGPQG IWGQCK-OH & \numrange{0.4}{4.2} & This study\\  \bottomrule
\end{tabularx}
\caption{Kinetic constant $k_\mathrm{cat}$ of MMP-1 enzyme reported in the literature with different substrates.}
\label{tab:kcat}
\end{table}

\newpage
\section{Erosion of parallel channels}
\label{sec:sup_parallel_channels}

\begin{figure}[ht]
\begin{center}
\includegraphics[width=1.0\textwidth]{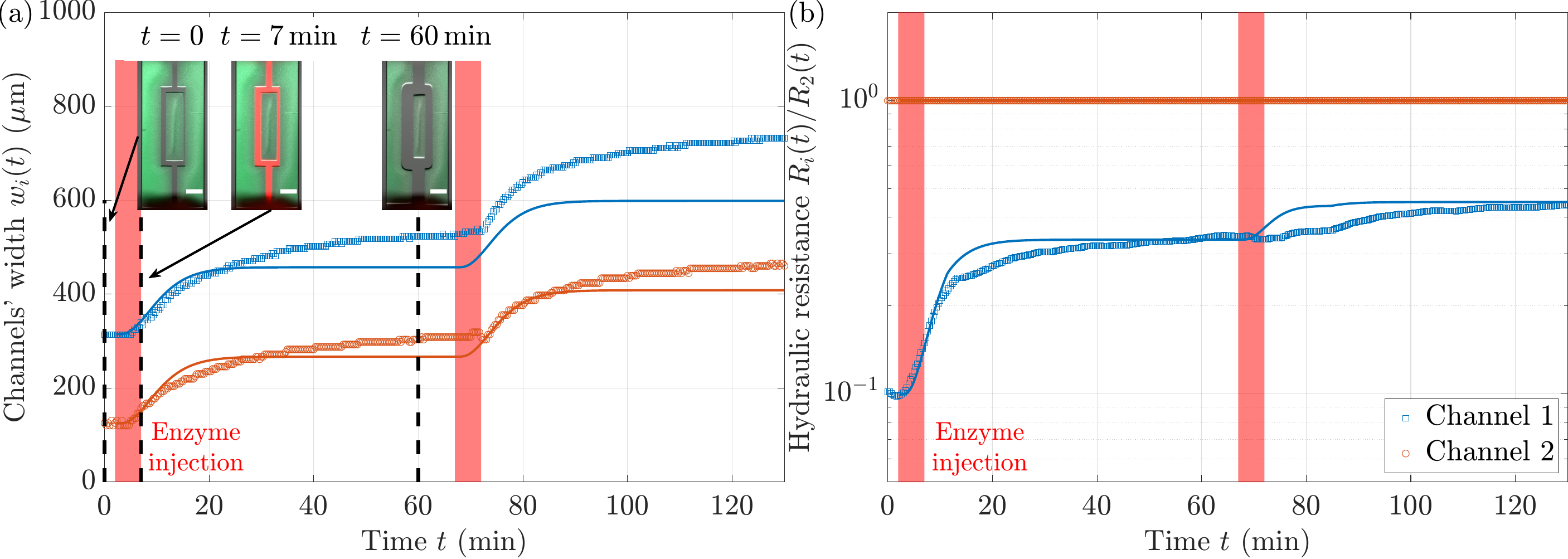}
\caption{\textbf{Erosion of two parallel channels.} (a) Temporal evolution of the channels' width in the two channel network, see Movie~2 in the Supplemental Material. Snapshots of the network at $t=0$, $t=7\,$min and $t=60\,$min are shown in inset. Channels are numbered from the widest to the narrowest. Points: experimental data. Lines: numerical simulations. Scale bar: 1\,mm. (b) Temporal evolution of the hydraulic resistance ratio $R_i(t)/R_2(t)$ for the same experiment. Points: experimental data. Lines: numerical simulations. The enzyme is injected for $t_e=5$\,min at $t=2\,$min and $t=\SI{67}{min}$ at a flow rate $Q=\SI{20}{\uL\per\min}$ (see Methods).}
\label{fig:width_2_channels}
\end{center}
\end{figure}

To study how erosion affects the homogenization of flow networks, the simplest imbalanced network was eroded, consisting of two parallel channels with identical height and length, but a different initial width (see Movie~2 in the Supplemental Material and Fig.~\ref{fig:width_2_channels}(a), inset). The erosion agent, the MMP-1 enzyme, was injected in two pulses of $t_e=5$\,min at a flow rate $Q=\SI{20}{\uL\per\min}$, separated by 60\,min of buffer injection at the same flow rate (see Methods). Initially, the wide 'Channel 1' and the narrow 'Channel 2' had respective widths $w_1=\SI{315\pm5}{\um}$ and $w_2=\SI{125\pm5}{\um}$. The width of the former increased by \SI{210}{\um} after each pulse, reaching a final width of \SI{735\pm5}{\um}. 'Channel 2' erosion was slightly slower, with a combined erosion of \SI{340}{\um}, to achieve a final width of \SI{465\pm5}{\um} [see Fig.~\ref{fig:width_2_channels}(a)].

Nevertheless, the widths eroded on the two channels are close. Consider the hydraulic resistances of the channels, $R_i = 12\eta L (H w_i^3)^{-1} (1-0.63w_i/H)^{-1}$ for a rectangular channel of length $L$, width $w_i$, height $H>w_i$ and fluid viscosity $\eta$~\cite{Bruus.2011}. There is an order of magnitude difference between the two channels before erosion [$R_1/R_2\simeq0.1$ at $t=0$, see Fig.~\ref{fig:width_2_channels}(b)]. The injection of the enzyme partly corrects this heterogeneity in the flow as the final ratio of hydraulic resistances at $t=135\,$min reaches $R_1/R_2\simeq0.43$. The erosion thus results in homogenizing the network.

\begin{figure}[ht]
\begin{center}
\includegraphics[width=1.0\textwidth]{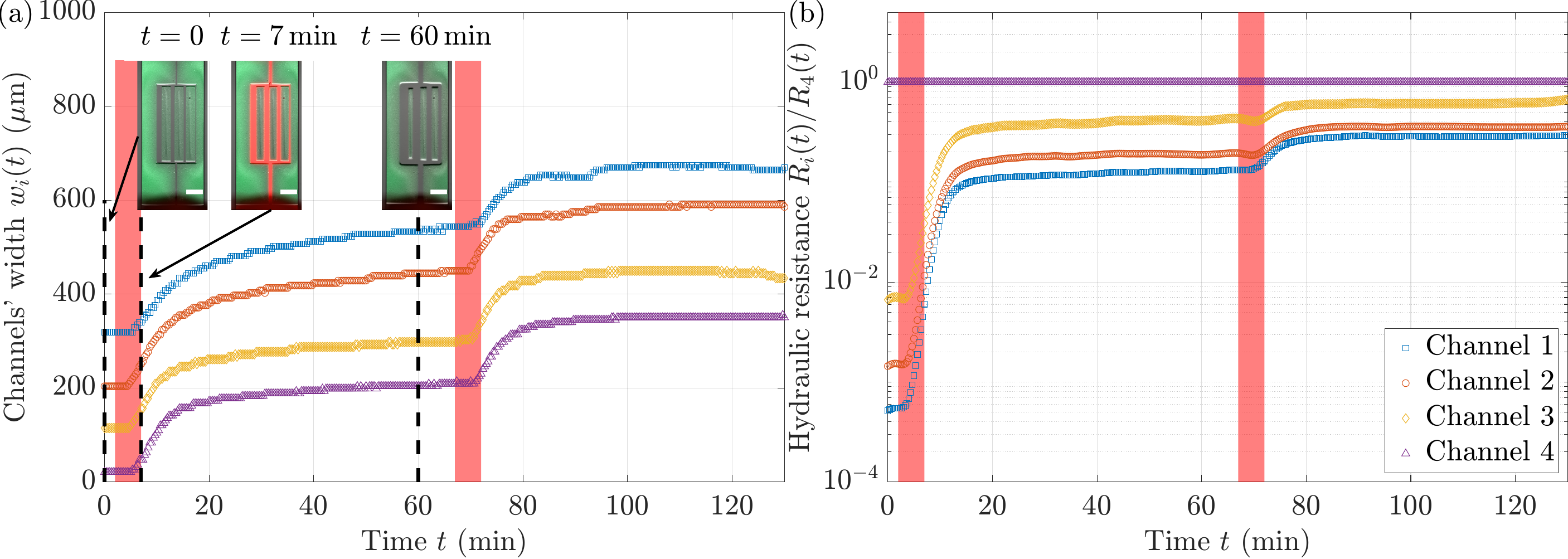}
\caption{\textbf{Erosion of four parallel channels.} (a) Temporal evolution of the channels' widths in the four-channel network (see Main~Fig.~1). Inset: snapshots of the network at $t=0$, $t=7\,$min and $t=60\,$min. Scale bar: 1\,mm. (b) Temporal evolution of the hydraulic resistance ratio $R_i/R_4$ for the same experiment. Channels are numbered from the widest to the narrowest. The enzyme is injected for $t_e=5$\,min at $t=2$\,min and $t=\SI{67}{min}$, at a flow rate of $\SI{20}{\uL\per\min}$; the same flow rate is used to inject  buffer before, in-between, and after the pulses. The  increase in width in all four channels is nearly identical, despite the different initial  channel widths.}
\label{fig:width_4_channels}
\end{center}
\end{figure}

A similar experiment was conducted in a four-channel network, see Main~Fig.~1, with initial widths ranging from $w_4=20$ to $w_1=\SI{320}{\um}$ (channels are numbered from the widest to the narrowest). We observed that the erosion rate of all parallel channels is the same, and the channel width increases by approximately the same amount in all channels, see Fig.~\ref{fig:width_4_channels}(a). After two pulses of the eroding agent, the width of the two smallest channels $w_3$ and $w_4$ increased by \SI{320}{\um}, while $w_2$ grew by \SI{380}{\um} and $w_1$ by \SI{355}{\um}. As with the two-channel network (see Fig.~\ref{fig:width_2_channels}), the impact of this similar erosion on the hydraulic resistances is significant. The resistance ratio $R_3/R_4$ increases from $\simeq\num{7e-3}$ at $t=0$ to 0.6 at $t=135\,$min. The resistance ratio $R_1/R_4$ between the widest and narrowest channels increases from $\simeq\num{5e-4}$ at $t=0$ to 0.3 at $t=135\,$min -- the accessibility of the least favorable path compared to the most favorable one increases more than 500-fold thanks to erosion. Erosion leads to tremendous homogenization of the network.

\newpage
\section{Enzyme concentration profile}
\begin{figure}[!b]
\begin{center}
\includegraphics[width=0.5\textwidth]{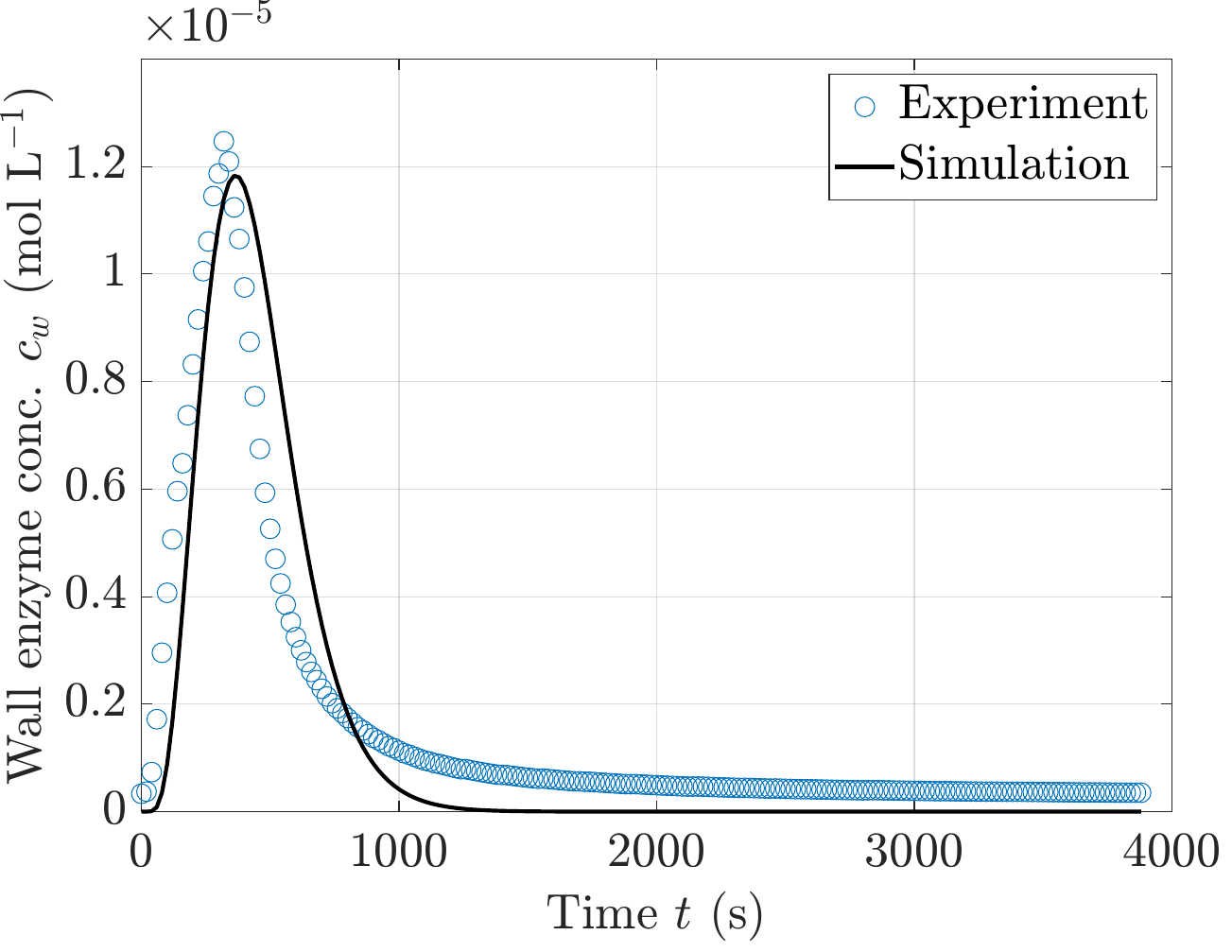}
\caption{\textbf{Enzyme concentration profile at the hydrogel wall.} Wall enzyme concentration $c_w$ for the single channel experiment, see Fig.~\ref{fig:erosion_rate}(a),(b), plotted for a single pulse of duration $t_e=\SI{5}{min}$. Experimental data: blue circles. The simulation data (black line), calculated as per Main~Eq.~(3b), is plotted with an absorption rate $K_a=\SI{8.4}{\um\per\s}$ and a desorption rate $K_d=\SI{11.4}{\per\s}$, all other parameters used in the simulation are the same as in the experiment.}
\label{fig:concentration}
\end{center}
\end{figure}

\clearpage
\section{Width homogenization in complex networks}
\label{sec:supmat_complex_networks}

\begin{figure}[ht]
\begin{center}
\includegraphics[trim = 0mm 0mm 0mm 0mm, clip, height=0.5\textwidth, angle=-90]{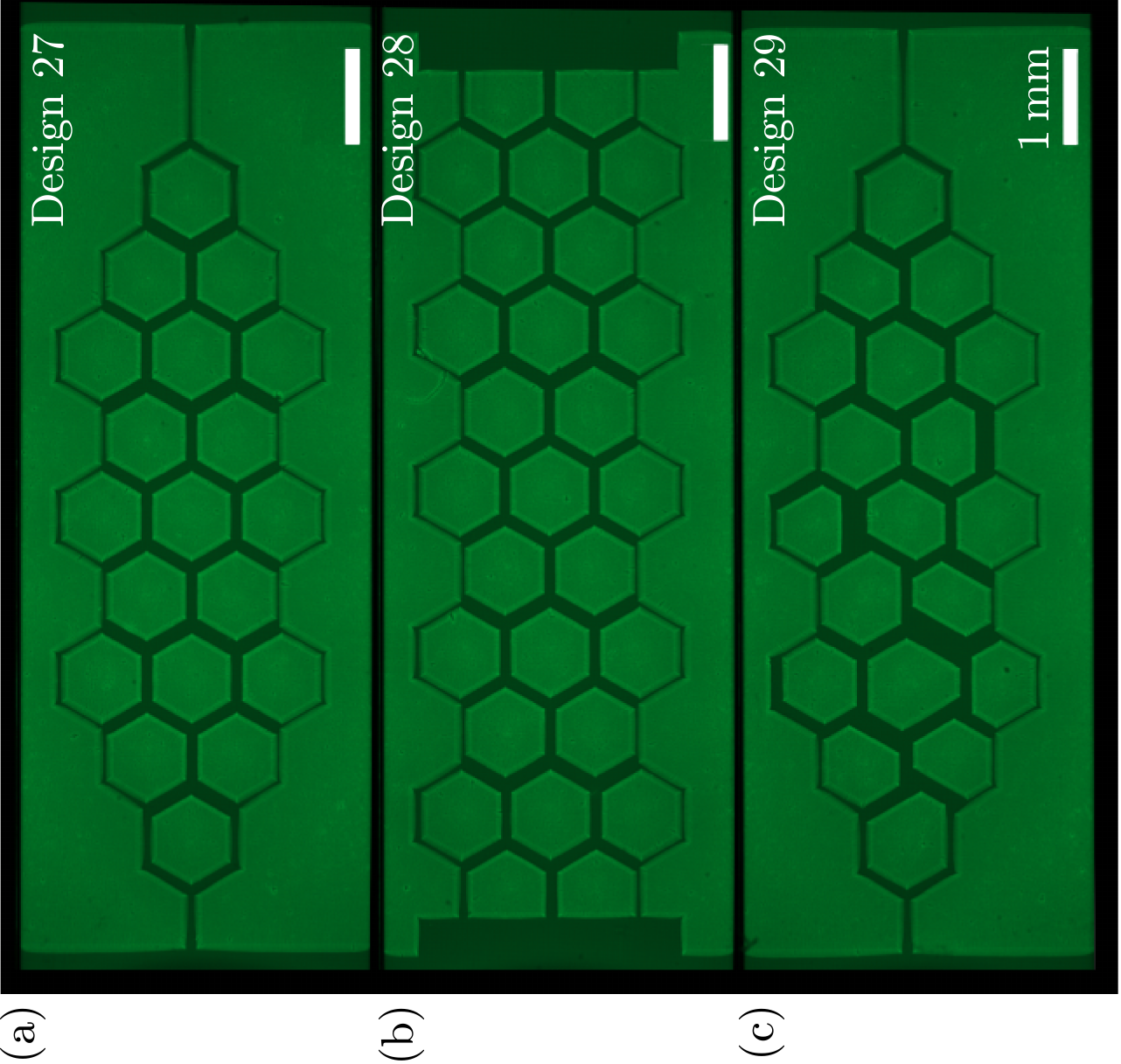}
\caption{\textbf{Design of the three different complex flow networks.}  (a) Design 27 is a hexagonal-based network, with wider channels in the center ($w\simeq\SI{135}{\um}$) and narrower channels on the sides ($w\simeq\SI{65}{\um}$). This is the design used for Main~Fig.~5. (b) Design 28 is similar to Design 27, but has three inlets and outlets instead of one each. (c) Design 29 is a variation of Design 27 with a randomization of the channel widths. Scale bar: 1\,mm.} 
\label{fig:hexagonalchannels}
\end{center}
\end{figure}

To test the robustness of homogenization by erosion, we study the homogenization of hexagonal networks. The three hexagonal networks tested each contain a couple dozens loops, see Fig.~\ref{fig:hexagonalchannels}. The first two, Fig.~\ref{fig:hexagonalchannels}(a),(b), are symmetric hexagonal loops of $\simeq\SI{550}{\um}$ side length, with respectively one and three in/outlets. The channels in the center are deliberately wider ($w\simeq\SI{135}{\um}$) than those on the sides ($w\simeq\SI{65}{\um}$) to see if the homogenization procedure is able to overcome those hurdles. Indeed, the hydraulic resistance on the side is higher because of both the width, which is smaller, and the length, which is longer (see Main~Eq.~(1)). The third network, Fig.~\ref{fig:hexagonalchannels}(c), is a variation of the first one with also only one in/outlet, but the widths of the channels have been randomized to mimic the randomness in living systems.

\begin{figure}[ht]
\begin{center}
\includegraphics[width=0.99\textwidth]{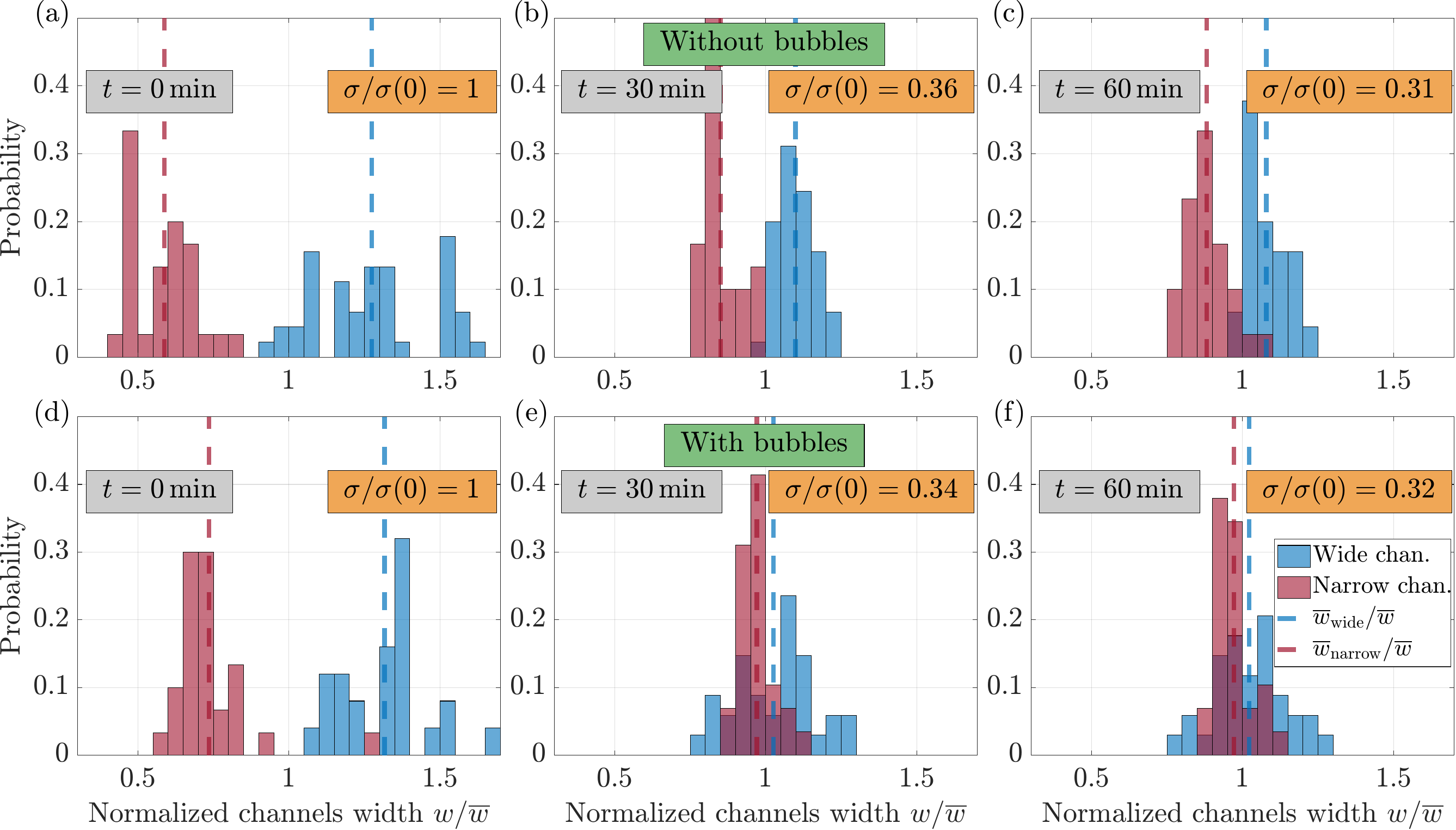}
\caption{\textbf{Homogenization of hydraulic resistances in Design 27.} Distribution of the channel widths $w$ normalized by the average width $\overline{w}$, for the symmetric hexagonal network Design 27. The initially wider (resp.~narrower) channels are plotted in blue (resp.~red). Distributions of channel widths for two experiments: one without any air bubbles (a, b \& c, see Movie~3 in the Supplemental Material), and one where bubbles are initially stuck in the wider channels of the network (d, e \& f, see Movie~4 in the Supplemental Material). The histograms showcase the start of the experiment ($t=0$, a \& d), after one enzyme pulse ($t=\SI{30}{min}$, b \& e) and just before the second enzyme pulse ($t=\SI{60}{min}$, c \& f). The injection protocol is the same as for the two-channel experiments: the enzyme is injected for 5\,min at $t=0$ and $t=\SI{65}{min}$ at a flow rate of $\SI{20}{\uL\per\min}$; the same flow rate is used to inject the buffer before, in-between, and after the pulses. Insets: normalized standard deviation $\sigma(t)/\sigma(0)$ of the normalized width of all the channels (wide and narrow).}
\label{fig:PDF_width_27}
\end{center}
\end{figure}

\begin{figure}[ht]
\begin{center}
\includegraphics[width=0.99\textwidth]{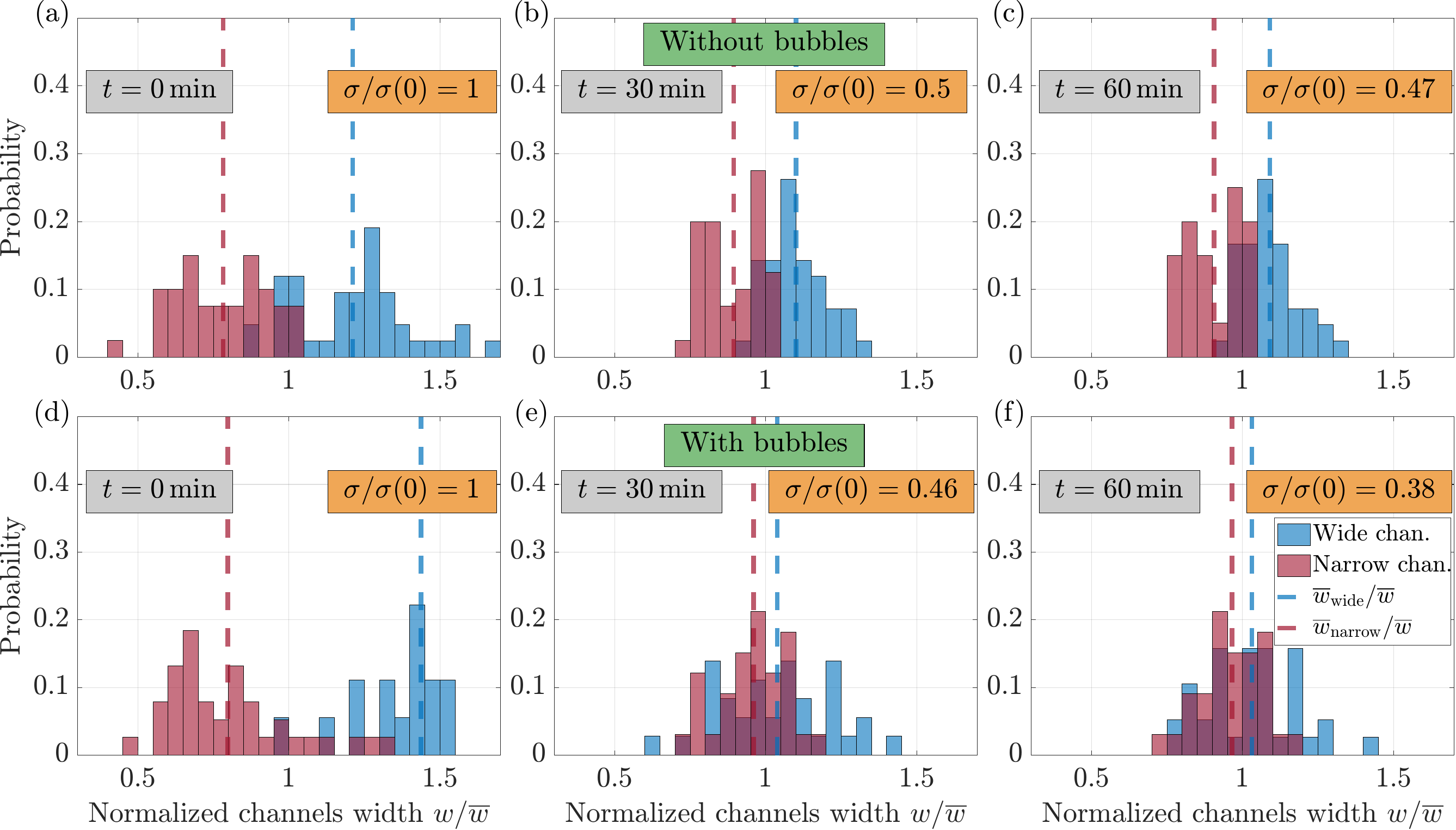}
\caption{\textbf{Homogenization of hydraulic resistances in Design 28.} Distribution of the channel widths $w$ normalized by the average width $\overline{w}$, for the symmetric honeycomb Design 28. Same as Fig.~\ref{fig:PDF_width_27}.}
\label{fig:PDF_width_28}
\end{center}
\end{figure}

The injection protocol follows the same procedure as described in Methods: the eroding agent is injected at a concentration \SI{3e-5}{\mol\per\liter} for $t_e=5$\,min at $t=0$ and $t=\SI{65}{min}$, at a flow rate of $\SI{20}{\uL\per\min}$; the same flow rate is used for the buffer which is injected before, in-between, and after the pulses (see Methods). The widths of all network channels are monitored during the experiments. For designs 27 \& 28, those segments are divided in two populations: the initially wide channels ($w\simeq\SI{135}{\um}$), and the initially narrow ones ($w\simeq\SI{65}{\um}$). Their width distributions, normalized by the global mean width, are plotted in Fig.~\ref{fig:PDF_width_27}(a)-(c) and Fig.~\ref{fig:PDF_width_28}(a)-(c) at different times of the experiment. In both cases, the results are similar: the initial distribution at $t=0$ is very wide, with the two channel populations clearly separated. 25\,min after the end of the first injection of enzyme, at $t=30\,$min, the widths get much closer even though the distribution is still bimodal (see Movie~3 in the Supplemental Material for Design 27). At $t=60\,$min, i.e.~right before the second pulse, the homogenization is even more pronounced. 

Although this erosion protocol largely homogenizes the channel widths, even in complex networks, it does not manage to reach a `perfect' homogenization state. Indeed, the widths of the narrow channels never entirely catch up with the widths of the wide channels. To optimize homogenization, we slightly change our protocol by initially injecting air bubbles in the network. Instead of carefully plugging the capillary tubing onto the Ibidi slide while the inlet pool is already full of liquid, we plug it while it is mostly empty. This allows the formation of air bubbles, which are then flowed into the network and obstruct most of the wide channels as their resistance is lower, see Movie~4 in the Supplemental Material, Fig.~\ref{fig:ColourCodedResistance} and Main~Fig.~5(e), redirecting the majority of the flow towards the narrower channels. The rest of the protocol is kept strictly identical, i.e.~we flow PBS at a flow rate $Q=\SI{20}{\uL\per\min}$ for 30\,min before injecting the eroding agent for $t_e=5$\,min at $Q$ and injecting again PBS for 60\,min at $Q$ (see Methods).

In the presence of air bubbles, the channel widths of the initially narrow channels catch up faster with the widths of the initially wide channels, leading to homogenization in the channel widths, see Main~Fig.~5(f). The widths distribution in
Fig.~\ref{fig:PDF_width_27}(d)-(f) and Fig.~\ref{fig:PDF_width_28}(d)-(f) show the same tendency with a much faster collapse into a single-peaked distribution. The bimodal characteristic of the channel width disappears, especially in Fig.~\ref{fig:PDF_width_27} for Design 27, highlighting the fact that the widths of the narrow population did catch up entirely. Thus, the initial injection of air bubbles does enhance the homogenization, by making it both quicker and better.

\begin{figure}[ht]
\begin{center}
\includegraphics[width=0.99\textwidth]{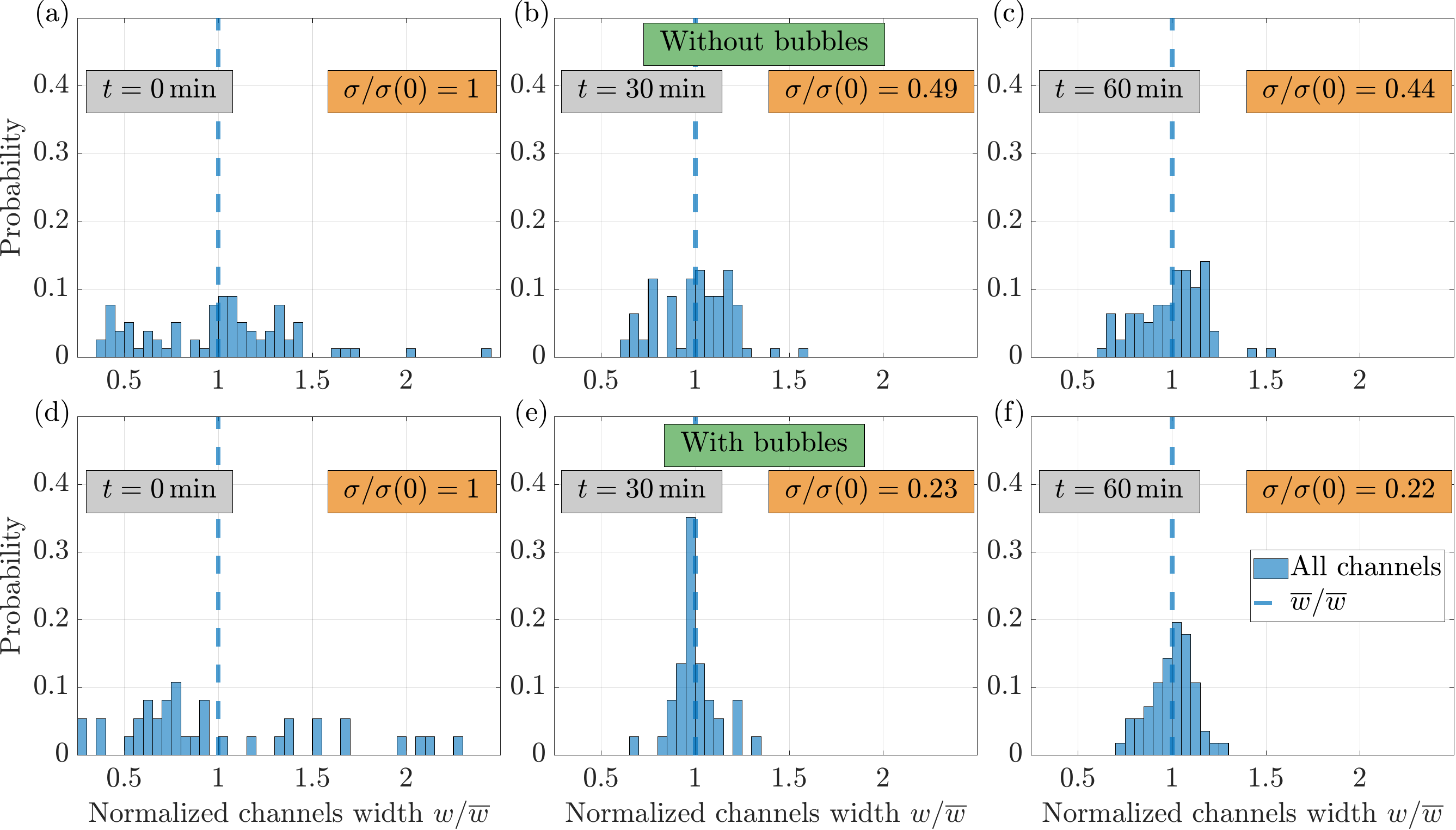}
\caption{\textbf{Homogenization of hydraulic resistances in Design 29.} Distribution of the channels' widths $w$ normalized by the average width $\overline{w}$, for the randomized honeycomb Design 29. Same as Fig.~\ref{fig:PDF_width_27}, except that the injection protocol for the experiment with bubbles is slightly different: The flow rate is ten times smaller, at $\SI{2}{\uL\per\min}$, and the enzyme injection time is $t_e=\SI{10}{min}$ (see Movie~5 in the Supplemental Material). This lower flow rate allowed the bubbles to stay longer in the network and have a more pronounced effect.}
\label{fig:PDF_width_29}
\end{center}
\end{figure}

For Design 29 [Fig.~\ref{fig:hexagonalchannels}(c)], the channels cannot be divided into two populations, as the initial channel widths are randomized. The same injection protocol as before, without air bubbles, is first used. The homogenization process still occurs in a similar fashion as with designs 27 \& 28, illustrated by the widths distribution in Fig.~\ref{fig:PDF_width_29}(a)-(c) which transitions from a widespread distribution to a more focused one. 

An additional experiment was conducted for Design 29 with air bubbles, with a slight change in the injection protocol: the flow rate was lowered to $\SI{2}{\uL\per\min}$ and the enzyme injection time was increased to 10\,min (see Movie~5 in the Supplemental Material). The widths distribution in Fig.~\ref{fig:PDF_width_29} shows an even stronger and faster transition to a peaked distribution of channel width, even though the comparison is less straightforward as the injection parameters were modified. The lowered flow rate allowed the air bubbles to stay longer in the microflow network. In the previous experiments at $\SI{20}{\uL\per\min}$, the bubbles were flushed away from the network as soon as they were no longer stuck. Here, with $Q=\SI{2}{\uL\per\min}$, the bubbles were not flushed immediately but instead moved from one channel to another. More specifically, they moved to the closest parts of the network that were newly eroded, following the path of least resistance. The bubbles were effectively acting as equalizers of the network by obstructing in real time the zones of low resistance, and thus slowing down the erosion in those parts.

\clearpage

\begin{figure}[h]
\begin{center}
\includegraphics[trim = 0mm 0mm 0mm 0mm, clip, width=0.99\textwidth, angle=0]{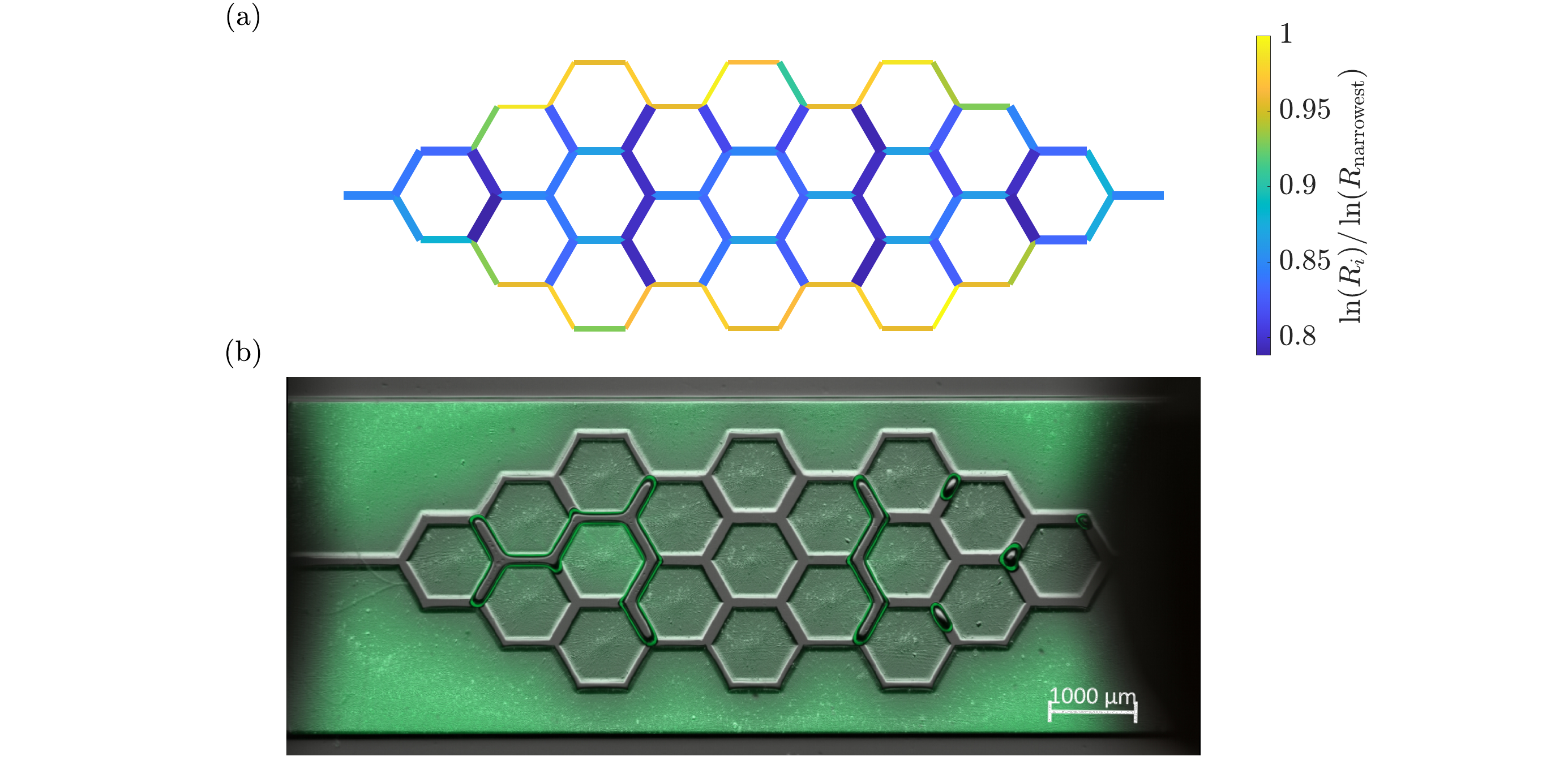}
\caption{\textbf{Air bubbles obstruct channels with the lowest resistances.}  (a) The normalized logarithms of the hydraulic resistances $\ln(R_i)/\ln(R_{\text{narrowest}})$ are color-coded at the start of the experiment shown in Main~Fig.~5(e), where $i$ stands for the different branches of the network. The branches in blue correspond to the lowest resistances whereas the yellow ones indicate the channels of highest resistances. The hydraulic resistances are calculated with Eq.~(1) from Main using the channel dimensions obtained from the experiment. (b) Snapshot of the experimental setup after the injection of the air bubbles. The bubbles are clearly located in the channels with the lowest resistances, i.e. $\ln(R_i)/\ln(R_{\text{narrowest}})<0.85$. Scale bar: 1\,mm.} 
\label{fig:ColourCodedResistance}
\end{center}
\end{figure}

\clearpage
\section{Erosion homogenizes the flow rates}

\begin{figure}[h]
\begin{center}
\includegraphics[width=0.5\textwidth]{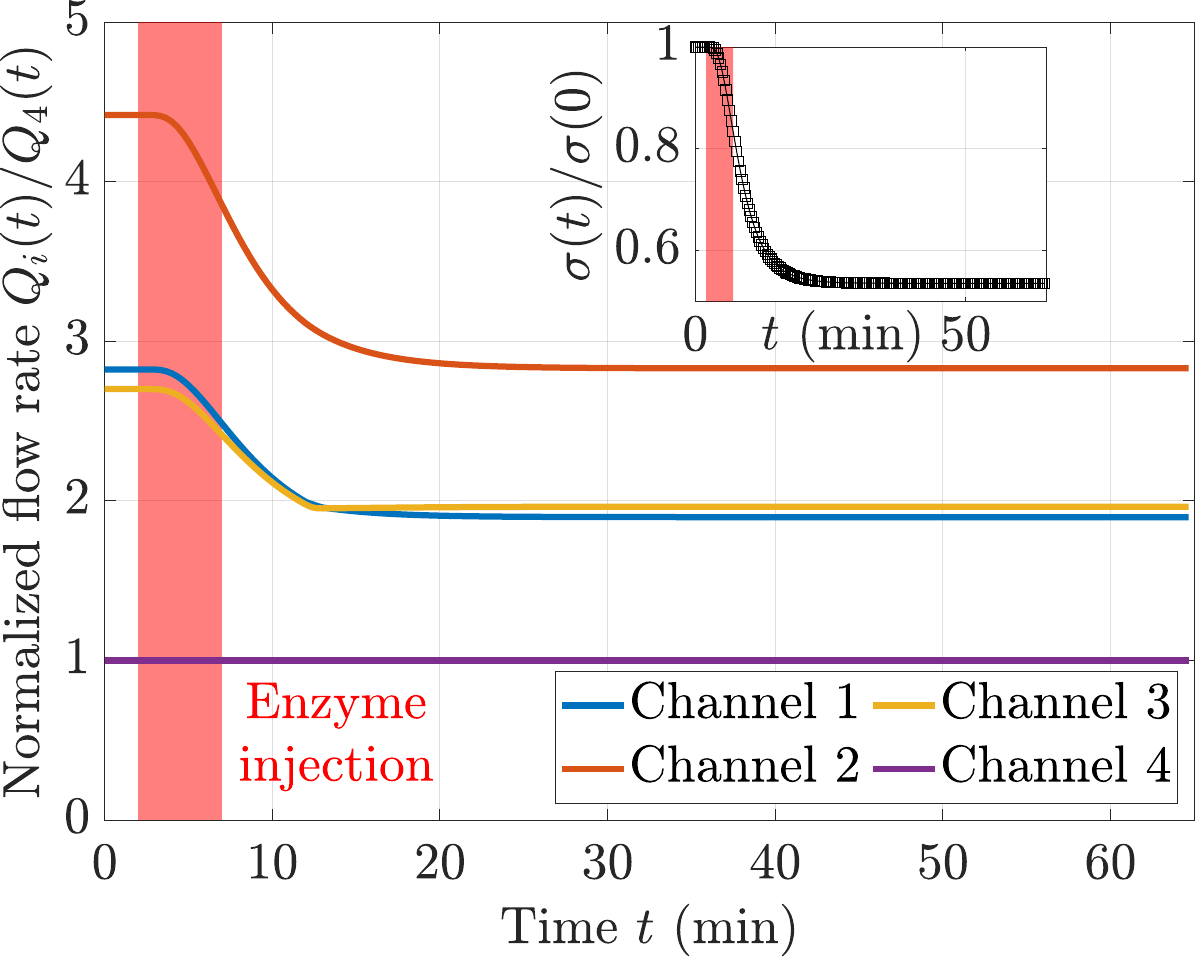}
\caption{\textbf{Erosion homogenizes the flow rates in the four-channel network.} Normalized flow rates $Q_i/Q_4$ obtained by simulating the experiment set up with 4-channels, where $i=1$ corresponds to the widest channel and $i=4$ to the narrowest channel, see Main~Fig.~1(a). The flow rate ratios homogenize as the channel walls erode. Inset: the normalized standard deviation $\sigma(t)/\sigma(0)$ of the normalized flow rates $Q_i(t)/Q_4(t)$ drops from 1 to 0.54, as a result of the flow rate homogenization thanks to erosion. Parameters used for the simulation were matched to the experimental parameters: inflow rate $Q=\SI{20}{\uL\per\min}$, diffusivity $D=\SI{30}{\square\um\per\s}$, absorption parameter $K_a=\SI{8.4}{\um\per\s}$, desorption parameter $K_d=\SI{11.4}{\per\s}$, channel height $H=\SI{400}{\um}$, enzyme concentration $c_0=\SI{3e-5}{\mol\per\liter}$, and erosion constant $\chi_0=\SI{3.3\pm0.6e3}{\um\per\min\mol\tothe{-1/2}\liter\tothe{1/2}}$. The widths of the channels at $t=0$ min were obtained from the experimental data.}
\label{fig:flowrate4channel}
\end{center}
\end{figure}

\begin{figure}[h]
\begin{center}
\includegraphics[width=0.99\textwidth]{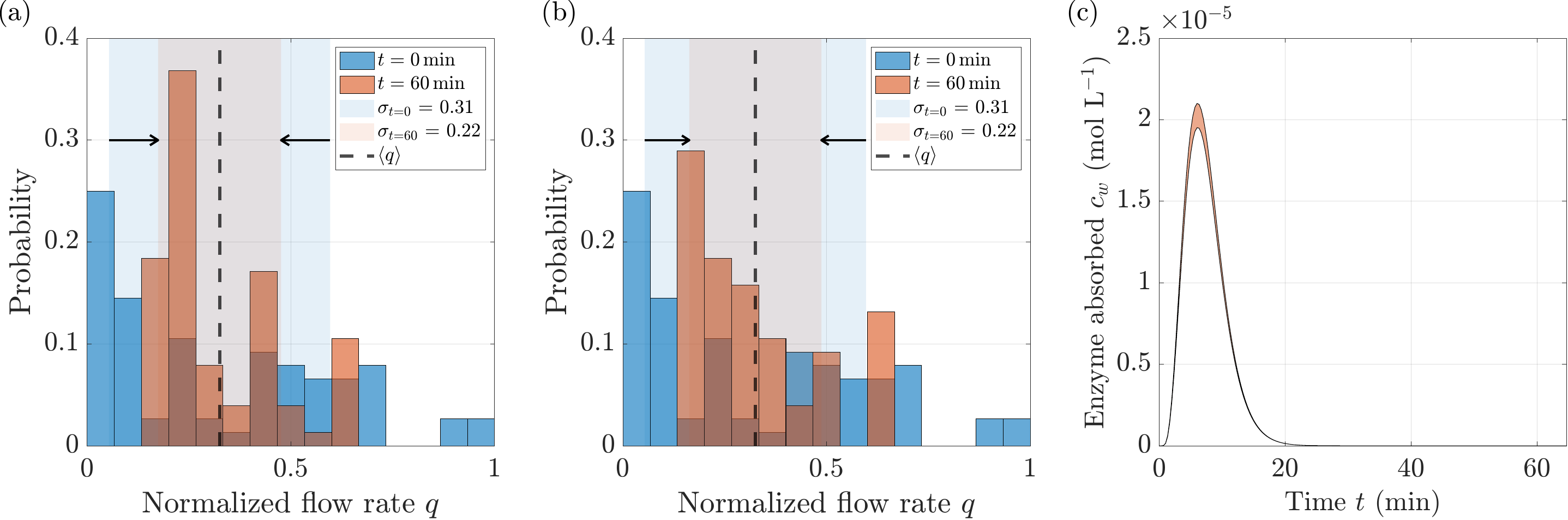}
\caption{\textbf{Erosion homogenizes the flow rates in the hexagonal network.} (a) Distribution of the normalized flow rates $q=Q/Q_{\rm max}$ before (in blue) and after erosion (in orange), in the complex network Design 27 [see Fig.~\ref{fig:hexagonalchannels}(a) and Main~Fig.~5(a)]. The channel widths used for calculating the flow rates were obtained from the experimental data. The standard distribution of the normalized flow rates drops from 0.31 before erosion (shaded in blue) to 0.22 after erosion (shaded in orange) as the flow rate distribution narrows, indicating homogenization of the flow rates. The average flow rate is plotted as a dashed line. (b) Flow rate distribution of the normalized flow rate obtained after simulating a pulse of enzyme through the  network Design 27 [see Fig.~\ref{fig:hexagonalchannels}(a) and Main~Fig.~5(a)]. Flow rates were calculated using the channel widths obtained through the simulation. The standard distribution of the normalized flow rates drops from 0.31 before erosion (shaded in blue) to 0.22 after erosion (shaded in orange), indicating homogenization of the flow rates. (c) Temporal evolution of the region between ${c_w}^{\rm max}(t)$ and ${c_w}^{\rm min}(t)$ (shaded in orange), where ${c_w}^{\rm max}(t)$ is the maximum value of $c_w$ across all the channels at time $t$ and ${c_w}^{\rm min}(t)$ is the minimum value of $c_w$ across all the channels at time $t$. The curve was obtained through the simulation shown in panel (b). A narrow shaded area shows that the enzyme concentration at the wall is similar across all the channels of the network at a given time $t$, leading to similar erosion in all the channels. Parameters used for the simulation were matched to the experimental parameters: inflow rate $Q=\SI{20}{\uL\per\min}$, diffusivity $D=\SI{30}{\square\um\per\s}$, absorption parameter $K_a=\SI{8.4}{\um\per\s}$, desorption parameter $K_d=\SI{11.4}{\per\s}$, channel height $H=\SI{400}{\um}$, enzyme concentration $c_0=\SI{3e-5}{\mol\per\liter}$, and erosion constant $\chi_0=\SI{3.3\pm0.6e3}{\um\per\min\mol\tothe{-1/2}\liter\tothe{1/2}}$. The widths of the channels at $t=0$ min were obtained from the experiment data.}

\label{fig:flowratedisthex}
\end{center}
\end{figure}

\clearpage
\section{Homogenization phase space}
\begin{figure}[h]
\centering
    \includegraphics[width = 0.99\textwidth]{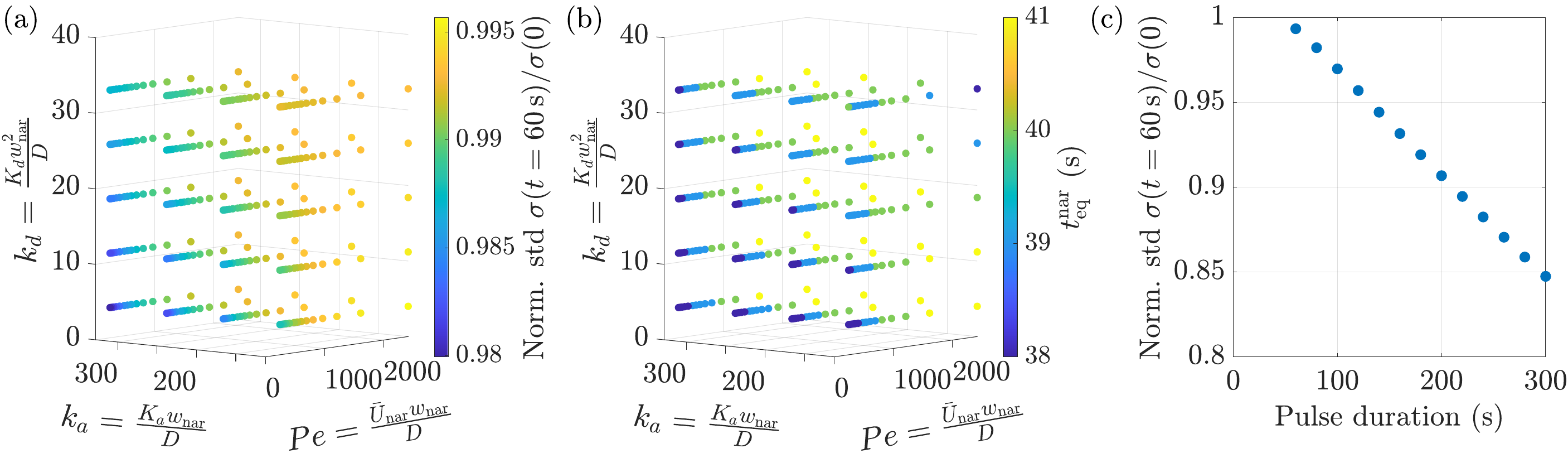}
  \caption{\textbf{Low Péclet, high absorption, and low desorption improve homogenization.}
  (a) The full three dimensional phase plot of the simulation results shown in Main~Fig.~3(c). See Movie~6 in the Supplementary Material for a rotating 3D view. (b) Time to equilibrate in the narrow channel $t_{\rm eq}^{\rm nar}$ (color coded) acts as a good predictor of the homogenization degree achieved through erosion. $t_{\rm eq}^{\rm nar}$ is lowest for high absorption, low desorption and low Péclet numbers. $D$, $K_a$ and $K_d$ varied for $w_{\mathrm{nar}}= \SI{200}{\um}$ and  $w_{\mathrm{wide}}= \SI{300}{\um}$. Parameters used: channel height $H=\SI{1}{mm}$, pulse length $t_e=\SI{1}{min}$ and inflow rate $Q= \SI{6}{\uL\per\min}$. (c) Increasing the pulse duration increases the amount of erosion, consequently improving homogenization. The simulations were performed with the same parameters as in Main~Fig.~3(c), by fixing $D=\SI{80}{\square\um\per\s}$, $k_a=80$, $k_d=16$, and varying the pulse duration from \SI{60}{\s} to \SI{300}{\s}.}
  \label{fig:theory_supp_3c}
\end{figure}


\section{Calculation of flow rates in the network edges}
 In a network, the flow rate $Q_{ij}$ in the edge connecting nodes $i$ and $j$ is given by $Q_{ij}=C_{ij}(P_i - P_j)$, where $C_{ij}$ is the inverse of the hydraulic resistance of the edge and $P_i$ is the pressure at node $i$. We calculate the flow rate at every edge of a network by imposing mass conservation at every node, known as Kirchhoff's law. Except for the inlet and outlet of the network (for which the inflow rates are known), every node has a net flow of zero. In general, net flow at a node $i$ is written as
$$Q_i = \sum_j C_{ij} (P_j - P_i),$$
where $j$ runs over all the nodes that are connected to $i$. The above equation for all the nodes is written succinctly in the matrix form 
\begin{align}
    \begin{pmatrix}
        \vdots \\
        Q_i \\
        \vdots   
    \end{pmatrix} = 
    \begin{pmatrix}
         & \cdots & \\
         \vdots & C_{ij} & \vdots \\
          & \cdots & \\
    \end{pmatrix}
    \begin{pmatrix}
        \vdots \\
        P_j \\
        \vdots   
    \end{pmatrix}
\end{align}
with
\begin{align*}
    Q_i :=& \text{ net flux at node $i$},\\
    P_j :=& \text{ hydraulic pressure at node $j$},\\
    C_{ij} :=& \text{ $1/R_{ij}$, where $R_{ij}$ is the hydraulic resistance of the edge connecting nodes $i$ and $j$},\\
    C_{ii} :=& -\sum_j C_{ij}, \text{ where $j$ runs over all the nodes connected to $i$.}
\end{align*}
Solving the above set of equations yields the pressures at each node $P_i$, which gives us the flow rate in the edge $ij$ connecting nodes $i$ and $j$ through the following relation,
\begin{align*}
    Q_{ij}=C_{ij} (P_j - P_i).
\end{align*}
Note that one needs to set one pressure value to zero to set a reference pressure.

\section{Effective advection diffusion coefficients with sorption (Zhang \emph{et al.})}

This section is a brief summary of the relevant results from Zhang \emph{et al.}~\cite{zhang_transient_2017} that have been used in our work. The goal is to model the transport of solute in a two-dimensional channel with sorbing boundaries, i.e.~the solute gets absorbed into the wall from the channel, while simultaneously getting desorbed out of the wall into the channel.
\subsection{Model}
An infinitely long two-dimensional channel where the flow is along the $x$-axis and channel width extends along the $y$-axis to $\pm w$ is assumed. The transport of a solute in such a channel can be modeled using the advection diffusion equation,
\begin{align}\label{AdvDiffDim}
    \pder[c]{t}=-U(Y)\pder[c]{x}+D\left(\pder[^2c]{x^2}+\pder[^2c]{y^2}\right),
\end{align}
where $c$ is the solute concentration [$N\,L^{-3}$], $t$ is time [T], $U(y)$ is the velocity profile of the fluid [$L\,T^{-1}$] and $D$ is the diffusion coefficient [$L^2\,T^{-1}$]. Owing to the symmetry of the problem, only the upper half of the channel is used for the calculations. The exchange of solute at the wall is modeled by the following boundary condition at the walls,
\begin{align}\label{BoundCond1Dim}
    -D\pder[c]{n}=\pder[C_w]{t},
\end{align}
where $C_w$ is the surface concentration of the absorbed solute [$N\,L^{-2}$] and $n$ is the direction along the normal to the surface. The boundary condition for sorbing boundaries assuming first order reaction kinetics at the wall is given by,
\begin{align}\label{BoundCond2Dim}
    \pder[C_w]{t}=K_ac-K_d C_w,
\end{align}
where $K_a$ is the absorption coefficient [$L\,T^{-1}$] and $K_d$ is the desorption coefficient [$T^{-1}$]. The quantities are non-dimensionalized as follows,
\begin{align}\label{NonDim}
    X = x/\xi, \quad Y= y/\xi, \quad u=U/\Tilde{U}, \quad
    C = c/C_0, \quad  C'_w= c_w/(C_0\xi), \quad T=t/(\xi^2/D),
\end{align}
where $\xi$ is the characteristic length scale (chosen to be the width of the channel $w$ in~\cite{zhang_transient_2017} and for the simulations in Main~Fig.~3, and height $H$ for other simulations), $\xi^2/D$ is the characteristic time scale, $\Tilde{U}$ is the cross-sectional velocity of the Poiseuille flow, $C_0$ and $C_0\xi$ are the characteristic concentration and surface concentration, respectively. After non-dimensionalizing \eqref{AdvDiffDim}, \eqref{BoundCond1Dim}, and \eqref{BoundCond2Dim}, we get
\begin{align}\label{AdvDiff}
    \pder[C]{T} = -\mathrm{Pe}\cdot u\pder[C]{X} + \pder[^2C]{X^2} + \pder[^2C]{Y^2},
\end{align}
with boundary conditions
\begin{align}\label{BoundCond}
    -\pder[C]{n}= \pder[C'_w]{T} = k_aC-k_d C'_w,
\end{align}
where we have a Péclet number: $\mathrm{Pe}=\frac{U_0 H}{D}$, and the two Damköhler numbers: $k_a=\frac{\xi K_a}{D}$ and $k_d=\frac{\xi^2K_d}{D}$.

\subsection{Solution}
The transverse averaged concentration $\bar{c}$, as introduced by Taylor (1953)~\cite{Taylor.1953}, is analyzed. The distribution of $\bar{c}$ along $x$ is described by its longitudinal moments $m_n=\int^{\infty}_{-\infty} x^n \bar{c} \diff{x}$, as introduced by Aris (1956)~\cite{Aris.1956}. It is assumed that the final cross-sectional averaged concentration profile along $x$ for a delta pulse at the channel entry point at $t=0$ can be approximated by the normalized lower order moments~\cite{zhang_transient_2017}, defined as,
\begin{align}\label{moments}
    M_0 = \frac{m_0}{m_I}, \qquad M_1=\frac{m_1}{m_0}, \qquad M_2=\frac{m_2}{m_0}-\left(\frac{m_1}{m_0}\right)^2.
\end{align}
where $m_I$ is the dimensionless initial mass. Thus the dimensionless velocity and longitudinal dispersion coefficient are given by,
\begin{align}\label{coefficients}
    v = \frac{dM_1}{dT}, \qquad D_L=\frac{1}{2}\frac{dM_2}{dT}.
\end{align}

The moments $m_n$ are given by~\cite{zhang_transient_2017},
\begin{align}
m_0(t)=& \sum_{k=0}^{\infty} a_k \exp \left(-p_k^2 t\right), \\
m_1(t)=& \sum_{k=0}^{\infty} b_k^{(1)} \exp \left(-p_k^2 t\right)+b_k^{(2)} t \exp \left(-p_k^2 t\right), \\
m_2(t)=& \sum_{k=0}^{\infty} c_k^{(1)} \exp \left(-p_k^2 t\right)+c_k^{(2)} t \exp \left(-p_k^2 t\right)+c_k^{(3)} t^2 \exp \left(-p_k^2 t\right),
\end{align}
where $p_k$'s are given by the solutions of~\cite{zhang_transient_2017},
\begin{align} \label{taneq}
    \tan(p)(p^2-k_d)-k_ap=0.
\end{align}
Each of the roots $p_k$ correspond to the characteristic decay rates of the moments with $p_0<p_1<p_2<\dots<p_\infty$. The analytical form of the coefficients $a_k$, $b^{(1)}_k$, $b^{(2)}_k$, $c^{(1)}_k$, $c^{(2)}_k$, and $c^{(3)}_k$ were derived by Zhang \emph{et al.}~\cite{zhang_transient_2017}. From the definition in \Eqref{moments}, we have,
\begin{align}
    M_1 = \frac{m_1}{m_0}&=\frac{\sum_{k=0}^{\infty} b_k^{(1)} \exp \left(-p_k^2 t\right)+b_k^{(2)} t \exp \left(-p_k^2 t\right)}{\sum_{k=0}^{\infty} a_k \exp \left(-p_k^2 t\right)}\\
    M_2= \frac{m_2}{m_0}-(M_1)^2 &= \frac{\sum_{k=0}^{\infty} c_k^{(1)} \exp \left(-p_k^2 t\right)+c_k^{(2)} t \exp \left(-p_k^2 t\right)+c_k^{(3)} t^2 \exp \left(-p_k^2 t\right)}{\sum_{k=0}^{\infty} a_k \exp \left(-p_k^2 t\right)} -(M_1)^2
\end{align}

Thus, from \Eqref{coefficients}, time derivatives of the normalized moments give the dimensionless effective advection coefficient $\Tilde{U}^{\mathrm{eff}}$ and diffusion coefficient $\Tilde{D}^{\mathrm{eff}}$, which are further related to the dimensional effective coefficients as ${U}^{\mathrm{eff}}=\Tilde{U}^{\mathrm{eff}}D/\xi$ and ${D}^{\mathrm{eff}}=\Tilde{D}^{\mathrm{eff}}D$. The zeroth order ($k=0$) terms determine the concentration profile at long time scales, and as we progressively include terms that are higher order in $k$, we get a more accurate description of the concentration profile at shorter time scales. For our work, we evaluate the moments by including terms up to $k=2$.

\section{Transport equation used in the simulations}
We define $\Bar{c}$ as the cross-sectional averaged enzyme concentration in the channel, $\Bar{N}$ as the number of enzyme particles in the channel, $C_w$ as the enzyme surface concentration at the wall, and $N_w$ as the number of enzyme particles in the wall. The channel has a width $w$, a height $H$ and a length $L$. Additionally, the channel walls are assumed to have a uniform thickness $\Delta h$ and $\Delta x$ is the smallest length unit of simulation. The equation which captures the time evolution of the enzyme concentration at the wall is:
\begin{align}\label{eq:wall_kin}
\pder[C_w]{t}=K_a\Bar{c}-K_d C_w.
\end{align}
In order to write down the above equation in terms of the number of particles, we convert the equation into volumetric concentrations by dividing both sides by a length scale $\Tilde{l}=\SI{1}{\milli\m}$. Then, assuming a wall thickness $\Delta h=\SI{1}{\um}$, we can write Eq.~\eqref{eq:wall_kin} as
\begin{align}\label{eq:wall_kin_num}
\pder[C_w]{t}\frac{(\Delta h \Delta x H)}{\Tilde{l}}&=K_a\bar{c}\frac{(\Delta h \Delta x H)}{\Tilde{l}}-K_d C_w \frac{(\Delta h \Delta x H)}{\Tilde{l}},\\
\pder[C_w]{t}(\alpha \Delta x H)&=K_a\bar{c}(\alpha \Delta x H)-K_d C_w (\alpha \Delta x H),
\end{align}
where we identify $\alpha=\Delta h/\Tilde{l}=10^{-3}$ as a non-dimensional parameter encoding the wall thickness relative to the length scale $\Tilde{l}$. Choosing $\Tilde{l}=\SI{1}{\milli\m}$ ensures that numerically $C_w$ has the same value as $c_w$, because the unit for length used in the simulations is $\SI{1}{\milli \m}$. Choosing $\Delta h=\SI{1}{\um}$ ensures that $\Delta h \gg \Delta w'=\chi_0\sqrt{c_0}\Delta t$, where $\Delta t$ is the simulation time step and $\Delta w'$ is the maximum erosion possible in time $\Delta t$ assuming the upstream concentration $c_0$ (for the maximum simulation time step used $\Delta t= \SI{0.01}{\s}$, $\Delta w'\sim \SI{0.003}{\um}\ll \Delta h$). Eq.~\eqref{eq:wall_kin_num} gives us the particle flux at the wall, the negative of which gives us the flux of particles in the channels (through the walls). Thus, for the number of particles in the channel, we get:
\begin{align}
\pder[\bar{c}(wH\Delta x)]{T}=-U^{\text{eff}}\pder[\bar{c}(wH\Delta x)]{x} + D^{\text{eff}}\pder[^2 \bar{c}(wH\Delta x)]{x^2} \underbrace{-K_a\bar{c}(\alpha \Delta x H)+K_d C_w (\alpha \Delta x H)}_{\text{wall flux}}
\end{align}
Dividing both sides by the channel volume, we can write the equation in terms of concentrations and see how the wall thickness plays an important role in the simulations.
\begin{align}
\pder[\bar{c}]{t}=-U^{\text{eff}}\pder[\bar{c}]{x} + D^{\text{eff}}\pder[^2 \bar{c}]{x^2} -\frac{\alpha K_a\bar{c}}{w} +\frac{\alpha K_d C_w}{w}.
\end{align}